\documentclass[11pt,a4paper]{article}
\pdfoutput=1
\usepackage{graphicx,epstopdf}
\usepackage{soul}
\usepackage{natbib}
\usepackage{amsmath,amsfonts,amssymb,amsbsy,amsthm,mathtools}
\usepackage{bbm,bm}
\usepackage{color,xcolor}
\usepackage{subfig}
\usepackage{booktabs}
\usepackage{url}
\usepackage{eurosym}
\usepackage[a4paper]{geometry}
\usepackage{a4wide}
\usepackage[colorlinks]{hyperref}
\usepackage{authblk}
\usepackage[linesnumbered,ruled,vlined]{algorithm2e}
\allowdisplaybreaks
\numberwithin{equation}{section}
\graphicspath{{Fig/}}
\usepackage{enumitem,float}
\usepackage[normalem]{ulem}

\newcommand{\R}{\mathbb{R}}
\renewcommand{\P}{\mathbb{P}}

\DontPrintSemicolon
\newcommand{\To}{\mbox{\upshape\bfseries to}}

\def\bX{{\boldsymbol{X}}}
\def\ba{{\boldsymbol{a}}}
\def\bb{{\boldsymbol{b}}}
\def\bx{{\boldsymbol{x}}}
\def\by{{\boldsymbol{y}}}

\def\bt{{\boldsymbol{t}}}
\def\bv{{\boldsymbol{v}}}

\def\bz{{\boldsymbol{z}}}

\newcommand{\bzero}{\bm{0}}

\def\H0{{\mathcal{H}_0}}

\def\btheta{{\boldsymbol{\theta}}}
\def\bvartheta{{\boldsymbol{\vartheta}}}
\def\bbeta{{\boldsymbol{\beta}}}
\def\bTheta{{\boldsymbol{\Theta}}}

\def\Lik{{\mathcal{L}}}
\def\cQ{{\mathcal{Q}}}
\def\cS{{\mathcal{S}}}
\def\DoA{{\mathcal{D}}}

\def\btheta{{\boldsymbol{\theta}}}
\def\beeta{{\boldsymbol{\eta}}}

\def\Qn{{\widetilde \cQ_n}}

\def\dif{\textrm{d}}

\def\dbeta{\textrm{Be}}
\def\MDA{\mathcal{D}}
\def\ntoinf{\stackrel{n\to\infty}{\longrightarrow}}
\def\ttoinf{\stackrel{t\to\infty}{\longrightarrow}}

\newcommand{\be}{\begin{eqnarray*}}
\newcommand{\ee}{\end{eqnarray*}}
\newcommand{\beq}{\begin{equation}}
\newcommand{\eeq}{\end{equation}}

\renewcommand{\leq}{\leqslant}
\renewcommand{\geq}{\geqslant}

\hypersetup{colorlinks,%
citecolor=blue,%
filecolor=green,%
linkcolor=red,%
urlcolor=violet,%
}

\newlist{inparaenum}{enumerate}{2}% allow two levels of nesting in an enumerate-like environment
\setlist[inparaenum,1]{label=(\roman*)}% labels for top level
\setlist[inparaenum,2]{label=(\roman{inparaenumi}\emph{\alph*})}% labels for second level

\title{Estimation and uncertainty quantification for\\ 
extreme quantile regions}
\author[1]{Boris Beranger}
\affil[1]{\footnotesize \emph{School of Mathematics and Statistics, University of New South Wales, Sydney}
\authorcr \emph{E-mail:} \url{B.Beranger@unsw.edu.au}, \url{Scott.Sisson@unsw.edu.au}}
\author[2]{Simone A. Padoan}
\affil[2]{\footnotesize \emph{Department of Decisions Sciences, Bocconi University} \authorcr
\emph{E-mail:} \url{Simone.Padoan@unibocconi.it}}
\author[1]{Scott A. Sisson}

\begin{document}

\maketitle

\begin{abstract}
Estimation of extreme quantile regions, spaces in which future extreme events can occur with a given low probability, even beyond the range of the observed data, is an important task in the analysis of extremes. Existing methods to estimate such regions are available, but do not provide any measures of estimation uncertainty. We develop  univariate and bivariate schemes for estimating extreme quantile regions under the Bayesian paradigm that outperforms existing approaches and provides natural measures of quantile region estimate uncertainty.
We examine the method's performance in controlled simulation studies. 
We illustrate the applicability of the proposed method by analysing high bivariate quantiles for pairs of pollutants, conditionally on different temperature gradations, recorded in Milan, Italy.
\end{abstract}


% =====================================================
% =====================================================
\section{Introduction}
\label{sec:Intro}
% =====================================================
% =====================================================

Estimating quantiles and how they can change depending on influential predictors (i.e.~quantile regression) is a recurring problem in many applied fields including medicine, survival analysis, economics, finance and environmental science  \citep[e.g.][]{yu2003quantile}. Assessing high quantiles, associated with fixed, low occurrence probabilities that lie beyond the range of $n$ existing data observations, is of crucial importance in risk management.
The solution to such a problem is not obvious when none of the observed data points  exceed this event. It is likely that the exceedance probability, $p$, is smaller than $1/n$. This is an extreme value problem.

More precisely, consider a random variable $X$ with distribution function $F$ defined on $\R_+:=(0,\infty)$. 
For $0<p<1$, let $Q(p):= F^{\leftarrow}(1-p)$ be the $(1-p)$-th quantile of $F$, where $F^{\leftarrow}$ is the left-continuous inverse
function of $F$, i.e.~$F^{\leftarrow}(x):=\inf\{y:F(y)\geq x\}$. Let $X_1,\ldots, X_n$ be independent random variables with distribution $F$. Interest is in estimating the quantile $Q(p)$ when the exceedance probability $p$ is very small.
We refer to $Q$ as an {\it extreme quantile}.
The extreme-value approach 
\citep[e.g.][]{BGST04, dHF06} assumes that as the sample size $n$ grows to infinity a suitable asymptotic probabilistic model can be used to approximate the desired quantile. In the asymptotic setting the exceedance probability depends on $n$, i.e.~$p=p_n$ with $p\to0$ as $n\to\infty$. Hence, for a sufficiently large number of observations, extreme-value theory can be used to compute
an approximation, $Q_n$, of the extreme quantile.  

A more challenging problem is estimating, for some fixed probability, an extreme bivariate region -- a subset of two-dimensional Euclidean space -- in which a future event would fall
when none of the two-dimensional observations fall in such a region,
and are likely to lie far from it \citep[e.g.][Ch. 6]{dHF06}.
In some applications there may not be any specific shape for the critical region, whereas in others it may be well defined. 
One illustration of the former is  air pollution monitoring, where the critical region is a set of combinations of two pollutants' concentrations where at least one of them is at a high level. Here, the shape of the critical region  depends on the type of pollutants and the intensity of their dependence.

\cite{EdHK2013} proposed a simple and practically useful method for defining critical regions which are generated through the level sets of a probability density function $f$, under suitable conditions (see Section \ref{ssec:biv_quantiles}). 
Precisely, a critical region is the level set of $f$ given by
\begin{equation}\label{eq:critical_reg}
\cQ=\{\bx\in\R^2_+:\,f(\bx)\leq \alpha\},
\end{equation}  
with $\alpha>0$, such that $\P(\cQ)=p$ for some very small $p\in(0,1)$. We refer to $\cQ$ as an {\it extreme quantile region}.
In particular $\cQ^{\complement}=\{\bx\in\R^2_+:\,f(\bx)> \alpha\}$ is the set with smallest area such that
$\P(\cQ^\complement)=1-p$. See also \cite{cooley2017} for an alternative approach.
Let $\bX_1,\ldots,\bX_n$ be independent two-dimensional random vectors with distribution function $F$. Similar to the univariate case, the extreme-value approach suggests adopting an appropriate asymptotic probabilistic model as $n$ approaches infinity. From this, a set $\cQ_n$, that is not fixed but depends on $n$, may be  derived such that $\P(\cQ_n)=p_n$ with $p\equiv p_n$ and $p\to 0$ as $n\to\infty$, which provides a sufficiently close approximation to the extreme quantile region.  
In particular, in both univariate and bivariate cases we assume the mild condition that $np\to c\in [0,\infty)$ as $n\to\infty$.

Beyond a definition and a method of estimation of such quantiles, a critical aspect in practice is to provide some quantification of the uncertainty around the estimates. Several estimators of extreme quantiles already exist in the univariate setting \citep[e.g.][]{BGST04, dHF06,  wang2015}. However, quantifying the uncertainty  of these estimators can be difficult
as their asymptotic variance depends on the parameters of second-order conditions which can 
be problematic to estimate in some settings. Estimators of extreme quantile regions also exist \citep{CEdH11, EdHK2013} but to the best of our knowledge  measures of their uncertainty are not available. A readily applicable uncertainty quantification for extreme quantiles is the methodological contribution of this article.

In this paper we develop a Bayesian approach for inferring both univariate extreme quantiles as well as extreme quantile regions. In the univariate case, we define a parametric Bayesian method for the extreme quantiles by exploiting the well known and widely used censored-likelihood approach \citep[e.g.][]{prescott1983, Smith94, LT96, huser2016, bienvenue2017},
 based on the likelihood function of the univariate Generalized Extreme Value (GEV) family of distributions \citep[e.g.][]{sisson2003}.
Inference is performed using an adaptive random-walk Metropolis-Hastings algorithm \citep{garthwaite2016}.
In the bivariate case we combine the univariate approach for the estimation of the GEV marginal parameters with the non-parametric Bayesian approach for the estimation of the extremal dependence proposed by \cite{MPAV2016}. 
As a result we obtain a semi-parametric Bayesian inferential method based on the censored-likelihood corresponding to a suitable two-dimensional extreme-value distribution. Through such a censored-likelihood it
is possible to simultaneously estimate the dependence structure of the extreme-value distribution together with the parameters of its margins, which are in turn members of the GEV class. 
The components of such a pseudo-posterior distribution are combined to produce a pseudo-posterior distribution for the extreme quantile regions.
Accordingly this approach allows for the direct estimation of extreme quantile intervals and quantile regions, as well as clear measures of their uncertainty.

There are many settings in environmental science
where the analysis of univariate extreme quantiles is of particular concern. These include evaluating tropical cyclone wind speed conditional on certain climate variables \citep{jagger2009modeling}, precipitation conditional on global climate model projections \citep{friederichs2012}, both precipitation and wind power given outputs from numerical weather prediction models \citep{bjornar2004, bremnes2004prob}, and ozone and particulate matter concentrations conditional on meteorological variables \citep{porter2015investigating}. 
However it is clear that some environmental variables depend on others: e.g.~wind speed is dependent on wind gust or air pressure \citep[e.g.][]{marcon2017semi}, and ozone is dependent on particular matter \citep[e.g.][]{falk2018}. 
Similarly, when designing offshore structures, engineers need estimates of extreme quantiles of oceanographic variables. Due the strong dependence between the involved variables e.g.~wave heights and wind speeds, bivariate extreme value tools (in the simplest case) are sometimes necessary \citep{naess2015statistics}.
As a consequence, the joint study of two (or more) dependent variables  will produce a more comprehensive and accurate analysis because it is based on more available information.

The goal of this article is to provide an inferential framework (i.e.~a method and  software to implement it)
 that can provide estimates of  extreme bivariate quantile regions, which can then be used to inform part of a larger study.
 For example, reliability engineers designing offshore platforms can use this framework as both a visual and quantitative tool to produce extreme quantile curves (such as those in Figure \ref{fig:biv_data_quant}) that can be informative for reliability design.
Here, we particularly focus on the problem of air pollution.

According to the World Health Organisations (WHO) and the Global Ambient Air Quality Database (2018 update, \url{https://www.who.int/airpollution/en/}), air pollution kills an estimated 7 million people worldwide each year, $4.2$ million of which as a result of exposure to ambient (outdoor) air pollution. Furthermore, $25\%$ of all heart diseases deaths are attributable to air pollution. 
Based on WHO guidelines \citep{world2006air} the European policy on air-quality standards \citep{guerreiro2016air} regulates emissions of the pollutants particulate matter (PM$_{10}$), ozone (O$_3$), nitrogen oxide (NO), nitrogen dioxide (NO$_2$) and sulphur dioxide (SO$_2$), aiming to reduce the negative impact that they have on human health, the environment and climate. 
For example, short-term pollutant concentrations (obtained through the empirical quantiles of pollutants time series) that should not be exceeded in order to protect against to air pollution peaks are: a daily average concentration of $50\,\mu$g/m$^3$ ($90.4$ percentile) for PM$_{10}$, a daily maximum concentration of $120\,\mu$g/m$^3$ ($93.2$ percentile) for O$_3$, an hourly average concentration of $200\,\mu$g/m$^3$ ($99.8$ percentile) for NO$_2$ and a daily average concentration of $125\,\mu$g/m$^3$ ($99.2$ percentile) for SO$_2$.
We refer to such concentrations as the pollutant limit thresholds.
In contrast, long-term pollutant thresholds are set based on yearly mean concentrations, rather than using quantiles. See \citet{guerreiro2016air} for details.  
The above thresholds are determined for each individual pollutant,  although it is well known that some pollutants are dependent on each other \citep[e.g.][]{dahlhaus2000graphical, clapp2001analysis, heffernan2004, world2006air}.

With our methodology (and software) we are able to estimate extreme quantiles while taking into account the dependence between pairs of pollutants, and thereby be informed about  potentially dangerous pollutant concentration combinations. 
This will be of particular interest to air pollution emission regulators and public health analysts.
Estimating extreme pollutant concentrations conditional on e.g.~meteorological variables can also provide information on whether some predictors are important for understanding the evolution of air pollution at high concentration levels.
Here, we investigate this explicitly through an analysis of the behaviour air pollution data recorded over the last $\sim20$ years in Milan, Italy. 
The remainder of this article is organised as follows. In Section \ref{sec:ext_quantiles} we briefly review the extreme-value approach for approximating the extreme quantiles in both the univariate (Section \ref{ssec:uni_quantiles}) and bivariate (Section \ref{ssec:biv_quantiles}) cases.  In Section \ref{S3:Inference} we
introduce our Bayesian semi-parametric approach for inferring extreme quantiles and extreme quantile regions.  Section \ref{sec:simu} provides an extensive simulation study examining the performance of the proposed method. The article concludes with an  analysis of the extremes of air pollution recorded in Milan (Section \ref{sec:data_analysis}) followed by a Discussion.

% =====================================================
% =====================================================
\section{Estimating extreme quantiles}
\label{sec:ext_quantiles}
% =====================================================
% =====================================================

% =====================================================
% =====================================================
\subsection{The univariate case}
\label{ssec:uni_quantiles}
% =====================================================
% =====================================================

Let $F$ be a distribution on $\R_+$, and assume that  
$F$ is in the domain of attraction of the GEV family of distributions, $F\in\DoA(G)$. Then, for $n=1,2,\ldots$ there are norming constants
$a_n>0$ and $b_n$ such that 
\begin{equation}\label{eq:GEV}
F^n(a_n x +b_n)\ntoinf\exp\left(-\left(1+\gamma x\right)_+^{-1/\gamma}\right)=:G(x;\gamma),
\end{equation}
where $\gamma>0$ is the extreme-value index that describes the heaviness of the tail of the distribution $G$ 
and $(a)_+=\max(a,0)$ \citep[see e.g.][Ch.~1]{dHF06}.
There are several extreme-value based methods for modelling extreme quantiles; see e.g.~\citet[Ch.~1, 4]{dHF06} and \citet{wang2015} for a compendium. Here, we briefly review one which is useful for our purposes.
By \citet[Theorem~1.1.6]{dHF06} the result in \eqref{eq:GEV} is also equivalent to the result
\begin{equation}\label{eq:exponent_function}
t\{1-F(a(t)x+U(t))\}\ttoinf \left(1+\gamma x\right)^{-1/\gamma},
\end{equation}
where $U(t) = F^{\leftarrow}(1 - 1/t)$ for $t>1$ and $a(\cdot)$ is a suitable function \citep[Theorem~1.1.6]{dHF06}.
Let $k\equiv k_n$ and assume that $k\to\infty$ and $k/n \to 0$ as $n\to\infty$.
Set $t=n/k$ and $y=a(n/k)x+U(n/k)$. Then, from \eqref{eq:exponent_function} we obtain
\begin{equation}\label{eq:GPD_approx}
F(y)\approx  1-\frac{k}{n}\left(1+\gamma \frac{y-\mu}{\sigma}\right)_+^{-1/\gamma},\quad n\to\infty,
\end{equation}
where $\mu,\sigma>0$ are location and scale parameters parameters. By few manipulations and \citet[Theorem~1.1.8]{dHF06} we have that $a(t) \approx \sigma$ and $U(t) \approx \mu$ as $t \to \infty$.
Result \eqref{eq:GEV} is also equivalent to 
\begin{equation}\label{eq:quant_conv}
\frac{U(tx)-U(t)}{a(t)}\ttoinf\frac{x^\gamma - 1}{\gamma},
\end{equation}
see \citet[Theorem~1.1.6][for details]{dHF06}.
Since $Q(p) \equiv U(1/p)$, from \eqref{eq:quant_conv} we obtain, using arguments in \citet[][Ch.~3.1]{dHF06},
\begin{align}\label{eq:extreme_q_approx}
Q(p) \approx \mu + \sigma \frac{\left( \frac{k}{np}\right)^\gamma -1}{\gamma}\quad\mbox{as } n\to\infty.
\end{align}

A suitable adjustment of the GEV distribution allows us to also derive the approximate quantile function in \eqref{eq:extreme_q_approx}.
Specifically, for some threshold $s>0$ and for $x\geq s$, let 
$F_s(x):=\P(X\leq x | X > s)$. 
Using \citet[Theorem~1.1.6]{dHF06}, for $a_n=a(n)$ and $b_n=U(n)$, for all $x\geq s$ we can write
\begin{equation}\label{eq:gpd}
F_{a_n s +b_n}(a_n x +b_n)\ntoinf 1- \left(1+\gamma \frac{x-s}{\tilde{\sigma}}\right)_+^{-1/\gamma}=: H((x-s)/\tilde{\sigma};\gamma),
\end{equation}
where $\tilde{\sigma}=1+\gamma s$ and $H(\cdot;\gamma)$ is the Generalized Pareto (GP) family of distributions. Set $y= a_n x + b_n$ and $t:= a_n s + b_n$, then for large $t$ and any $y\geq t$ 
we obtain
\begin{eqnarray}
\nonumber F(y)&\approx& F(t)+\{1-F(t)\}H\left((y-t)/\bar{\sigma};\gamma\right)
\nonumber\approx \exp\left[-\{1-F(t)\}\left\{1-H\left((y-t)/\bar{\sigma};\gamma\right)\right\}\right]\\
\nonumber&\approx& \exp\left[-t^{-1}\left\{1-H\left((t-\mu)/\sigma;\gamma\right)\right\}\left\{1-H\left((y-t)/\bar{\sigma};\gamma\right)\right\}\right]\\
\label{eq:gev_app}&=& \exp\left(-t^{-1}\left(1+\gamma (y-\mu)/\sigma\right)_+^{-1/\gamma}\right)\equiv G^{1/t}((y-\mu)/\sigma;\gamma),
\end{eqnarray}
where $\bar{\sigma}=\sigma+\gamma(t-\mu)$. 
Writing again $F(y)=1-p$ and $t=n/k$ in \eqref{eq:gev_app}, and by noting that
$-\log(1-p)\approx p$ for $p\to 0$, then the quantile function in \eqref{eq:extreme_q_approx} can be obtained by inverting the expression in \eqref{eq:gev_app} with respect to $p$.

% =====================================================
% =====================================================
\subsection{The bivariate case}
\label{ssec:biv_quantiles}
% =====================================================
% =====================================================

Let $\bX=(X_1,X_2)$ be a random vector with joint distribution function $F$ on $\R^2_+$ with
margins $F_j$, $j=1,2$, and probability density function $f$.
Assuming that $F\in\MDA(G)$, then there are sequences of norming constants $\ba_n>\bzero$ and $\bb_n$ such that
\begin{equation}\label{eq:multi_GEV}
F^n(\ba_n\bx+\bb_n)\ntoinf G(\bx),
\end{equation}
where $G$ is the so-called bivariate max-stable distribution \citep[Ch~6]{dHF06}, whose margins 
$G(x_j;\gamma_j)$ are members of the GEV family  \eqref{eq:GEV} with tail indices $\gamma_j>0$, $j=1,2$.
In the following we denote
$
G_*(\bx):=G(x_1^{\gamma_1},x_2^{\gamma_2}),\quad \bx\in \R_+^2,
$
as an extreme-value distribution with
unit-Fr\'echet margins: $G_*(x_1,\infty)=\exp(-1/x_1)$ and $G_*(\infty,x_2)=\exp(-1/x_2)$ for
every $x_1,x_2>0$.

The convergence result in \eqref{eq:multi_GEV} implies convergence at both marginal and dependence levels.
For marginal convergence we have that \eqref{ssec:uni_quantiles}, \eqref{eq:exponent_function} and \eqref{eq:GPD_approx} hold for
$F_j$, with $a_{n,j}$, $b_{n,j}$, $a_j(t)$, $U_j(t)$ and  $y_j=a_j(n/k)x_j+U_j(n/k)$ with $j=1,2$. Furthermore, $a_j(t) \approx \sigma_j>0$ and $U_j(t) \approx \mu_j>0$ as $t \to \infty$.
For convergence of the dependence structure, for every $(x,y)\in (0,\infty]^2\setminus \{(\infty,\infty)\}$, from \citet[Ch~6.1.2]{dHF06} we have that
\begin{eqnarray}\label{E:DoA_df}
t\left(1-F(tU_1(x_1), tU_2(x_2))\right)
\ttoinf -\log G_*(\bx).
\end{eqnarray}
The result in \eqref{E:DoA_df} implies the existence of a measure $\nu$, named the exponent measure \citep[see][Ch~6 for details]{dHF06} such that for every $\bx\in \R_+^2$,
\begin{equation*}\label{eq:exp_fun}
-\log G_*(\bx)=\nu(\{\bv\in \R^2_+: v_1>x_1 \text{ or } v_2>x_2\})=2\int_0^{1}\left(\frac{w}{x_1}\vee \frac{1-w}{x_2}\right)H(\dif w),
\end{equation*}
where $H$ is a probability measure on $[0,1]$ satisfying the mean constraint
$
\int_0^{1} w H(\dif w)= \int_0^{1}(1-w) H(\dif w)=1/2.
$
In the following we denote both the probability measure and its distribution function by $H$, with the difference being determined by the context. 

The extreme quantile regions method introduced in \cite{EdHK2013} requires also some further assumptions at the density level.
The density $f$ is assumed to be decreasing in each coordinate, outside of $(0,M]^2$ for some $M>0$, and
bounded away from zero on $(0,M]^2$. 
There exist a nonnegative Lebesgue integrable function $g$ such that for every $\bx\in \R_+^2$, 
$$ 
\nu(\{\bv\in \R^2_+:v_1>x_1 \mbox{ or } v_2>x_2\})=\iint_{\{v_1>x_1 \mbox{ or } v_2>x_2\}} g(\bv) \dif \bv.
$$
and on $\R^2_+$,
\begin{align}\label{eq:DoA_density}
tU_1(t)U_2(t)f\left(tU_1(x_1), tU_2(x_2)\right)
\ttoinf (\gamma_1\gamma_2)^{-1}x_1^{1-\gamma_1}x_2^{1-\gamma_2}g(\bx) =: q(\bx).
\end{align}
We refer to $g$ and $q$ as the density of the exponent measure the {\it basic density} function, respectively.
By the above assumption on $g$ it follows that $H$ has a continuous
density $h:=\partial H/\partial w $ on $(0,1)$ with no atoms at $0$ and $1$, i.e. $H(\{0\})=H(\{1\})=0$.
By the homogeneity of $g$ we have that $h(w)=2^{-1}g(w, 1-w)$, where  $r=x_1+x_2$ and $w=x_1/r$. 
We refer to $H$ and $h$ as the angular measure and density, respectively. 

The extreme-value approach for modelling extreme quantile regions $\cQ$ in \eqref{eq:critical_reg},
works with the set $\cQ_n=\left\{\bx\in \R^2\colon f(\bx)\leq\alpha\right\}$,  where $\alpha=\alpha_n$ is not fixed but depends on the sample size $n$ such that $\P(\cQ_n)=p$ with $p=p_n\to 0$ as $n\to\infty$.
In particular, \cite{EdHK2013} suggest focusing on the fixed set $\cS=\{\bx\in \R^2_+: q(\bx)\leq 1\}=\left\{\bx\in \R^2_+\colon r\geq q_*^{-1}(w),w\in[0,1]\right\}$
that we call the {\it basic set}, where 
\begin{equation}
\label{eq:dens_exp_measure}
q_*(w) = q(w,1-w)^{-\frac{1}{{1+\gamma_1+\gamma_2}}}, \quad
q(w,1-w)=\frac{2w^{1-\gamma_1}(1-w)^{1-\gamma_2}h(w)}{\gamma_1\gamma_2}.
\end{equation}
and where $q(w,1-w)$ is obtained from the relation $q(rw,r(1-w))=q(w,1-w)/r^{(1+\gamma_1+\gamma_2)}$.
Abusing terminology refer to $q_*$ as the {\it angular basic density} function.
According to the exponent measure, the size of $\cS$ is
\begin{align}\label{eq:measure_basic_set}
\nu(\cS)=2\int_{[0,1]}q_*^{-1}(w)\,h(w)\dif w.
\end{align}
The idea is to then inflate the basic set $\cS$ 
into an extreme set, $\Qn$, depending on $n$, so that $\P(\Qn)\approx p$ for $n\to\infty$
and such that $\Qn$ is a good approximation of $\cQ_n$, i.e.~so that $\P(\cQ_n \triangle \Qn)/p\to 0$ as
$n\to\infty$, where $B\triangle D = B \backslash D \,\cup\, D\backslash B$, for two nonempty sets $B$
and $D$.
Here we consider a slightly different definition of the set $\Qn$ to that given in \cite{EdHK2013}. 
Specifically, by exploiting the fact that $U_j(n/k)\approx\mu_j$ and $a_j(n/k)\approx\sigma_j$, $j=1,2$ as $n\to\infty$, then for large $n$, we have that $\Qn$ can be approximated as 
\begin{equation}\label{eq:Q_n_approx}
\Qn\approx \left\{\left(\mu_1 + \sigma_1\frac{\left(\frac{k\nu(\cS)x_1}{np}\right)^{\gamma_1}-1}{\gamma_1},  
\mu_2 +\sigma_2\frac{
\left(\frac{k\nu(\cS)x_2}{np}\right)^{\gamma_2}-1}{\gamma_2}\right)\colon (x_1,x_2)\in \cS\right\}.
\end{equation}
The approximation in \eqref{eq:Q_n_approx} is consistent with the formula in 
\eqref{eq:extreme_q_approx} used to approximate the univariate extreme quantiles. In Section \ref{ssec:infer_biv} we show how the bivariate max-stable distribution in \eqref{eq:multi_GEV} can be used to
estimate the extreme quantile region in \eqref{eq:Q_n_approx}.

% =====================================================
% =====================================================
\section{Inference}
\label{S3:Inference}
% =====================================================
% =====================================================

% =====================================================
% =====================================================
\subsection{The univariate case}
\label{ssec:infer_univ}
% =====================================================
% =====================================================

We describe an approximate Bayesian framework for estimating the extreme quantile in \eqref{eq:extreme_q_approx}
for small $p$ (where the meaning of ``small $p$'' is given in Section \ref{sec:Intro}). 
In particular, we explore the Bayesian paradigm using a censored likelihood  \citep{sisson2003},
based on \eqref{eq:gev_app}.
Specifically, let $X_1,\ldots, X_n$ be independent and identically distributed random variables with distribution function $F$ on $\R_+$, where $F\in\DoA(G)$.
First we define a high threshold $t$.
Let $X_{1,n}\leq X_{2,n}\leq\cdots\leq X_{n,n}$ be the $n$ order statistics and $F_n$ be the
empirical distribution function. The threshold may then be defined as $T=X_{n-k,n}$ for large $k$ such
that $1-F_n(X_{n-k,n})=k/n$ is close to zero, for instance $k/n=0.10, 0.05,0.01$.
Next, let $\by_{1:n}=(y_1,\ldots,y_n)$ be a realisation of $(X_1,\ldots, X_n)$ and
$t$ be the corresponding threshold. Then on the basis of the approximation in 
\eqref{eq:gev_app} we define the censored likelihood function
$$
\Lik(\by_{1:n};\btheta)=\prod_{i=1}^n\Lik(y_i;\btheta),
$$
where each contribution to the likelihood depends on the domain where an observation $y_i$ falls. That is, 
\begin{equation}
\label{eq:uni_CensLik}
\Lik(y_i;\btheta) \propto
\begin{cases}
G^{k/n}(t;\btheta), & \mbox{ if } y_i\leq t,\\[1em]
\frac{\partial}{\partial y} G^{k/n}(y;\btheta)|_{y=y_i}, & \mbox{ if } y_i>t,\\[1em]
\end{cases}
\end{equation}
where $G^{k/n}(y;\btheta)\equiv G^{k/n}((y-\mu)/\sigma;\gamma)$ with $\btheta=(\mu, \sigma,\gamma)^\top$, $G^{k/n}((y-\mu)/\sigma;\gamma)$ is given in \eqref{eq:gev_app} and
$$
\frac{\partial}{\partial y} G^{k/n}(y;\btheta)|_{y=y_i} = G^{k/n}((y_i-\mu)/\sigma;\gamma)\left(1+\frac{\gamma(y_i-\mu)}{\sigma}\right)^{-1/\gamma-1} \frac{1}{\sigma} \frac{k}{n}.
$$

Assuming a prior $\Pi(\btheta)$ for $\btheta\in\bTheta\subseteq\R_+^3$, we draw samples from the resulting posterior distribution using an adaptive Metropolis-Hastings algorithm.
Specifically, we directly apply the adaptive (Gaussian) random-walk Metropolis-Hastings (RWMH) algorithm discussed in \citet{garthwaite2016}.
The current state of the chain $\btheta^{(j)}$ at time $j$ is updated by proposing a draw 
$\btheta'\sim h(\btheta| \btheta^{(j)})=\phi_3(\btheta^{(j)}, \tau^{(j)} \Sigma^{(j)})$ where $\phi_d(\ba, A)$ denotes a $d$-dimensional Gaussian density function with mean $\ba$ and covariance matrix $A$. 
Because  $h(\btheta'| \btheta) = h(\btheta | \btheta')$ is symmetric the acceptance probability of setting $\btheta^{(j+1)}=\btheta'$ reduces to
$$
\pi^{(j)} = \min \left( \frac{\Lik(\by_{1:n};\btheta')\Pi(\btheta')}{\Lik(\by_{1:n};\btheta^{(j)})\Pi(\btheta^{(j)})}, 1\right),
$$
otherwise $\btheta^{(j+1)}=\btheta^{(j)}$.
Following \citet{haario+st01}, the proposal covariance matrix $\Sigma^{(j)}$ is specified as
\begin{equation}
\label{eq:A_update}
\Sigma^{(j+1)} = \left\{
\begin{array}{ll}
(1 + [\tau^{(j)}]^2 / j) \mathrm{I}_3 , & j \leq 100 \\
\frac{1}{j-1} \sum_{k=1}^j (\btheta^{(k)} - \bar{\btheta}^{(j)})(\btheta^{(k)} - \bar{\btheta}^{(j)})^\top + ([\tau^{(j)}]^2/j) \mathrm{I}_3, & j > 100,
\end{array}
\right.
\end{equation}
where $\mathrm{I}_d$ is the $d$-dimensional identity matrix, $\bar{\btheta}^{(j)}=j^{-1}(\btheta^{(1)}+\cdots+\btheta^{(j)})$,
and $\tau^{(j)} >0$ is a scaling parameter that affects the 
acceptance rate of proposal parameter values.
Following \citet{garthwaite2016} we adaptively update $\tau$ using a Robbins-Monro process so that
\begin{equation}
	\label{eq:updatetau}
	\log \tau^{(j+1)} = \log \tau^{(j)} + c (\pi^{(j)} - \pi^*),
\end{equation}
where $c = (2\pi)^{1/2} \exp(\zeta_0^2/2) / (2\zeta_0)$ is  a steplength constant, $\zeta_0 = - 1/\Phi(\pi^*/2)$, and where $\Phi$ is the univariate standard Gaussian distribution function. The parameter $\pi^* $ is the desired overall sampler acceptance probability, here specified as $\pi^*= 0.234$ following \citet{roberts1997}.
This algorithm is summarised in Step 1 of Algorithm~\ref{alg:algo_joint}. 
Proposed Gaussian updates for the GEV scale parameter are performed on $\log\sigma$. See \citet{garthwaite2016} for further details.
% 

% =====================================================
% =====================================================
\subsection{The bivariate case}
\label{ssec:infer_biv}
% =====================================================
% =====================================================

Inference for the extreme quantile region $\Qn$ in \eqref{eq:Q_n_approx} requires
estimation of the two sets of marginal parameters $\btheta_i=(\mu_i, \sigma_j,\gamma_i)$, $i=1,2$, together with the basic set $\cS$ and its measure $\nu(\cS)$. In particular,
the estimation of the basic set $\cS$ its measure $\nu(\cS)$ in \eqref{eq:measure_basic_set} requires estimation of the angular density $h$. 
We extend the Bayesian procedure of the univariate case (Section \ref{ssec:infer_univ}) to the bivariate setting to  simultaneously estimate the marginal parameters and the angular density. This framework utilises the censored likelihood based on the bivariate max-stable distribution in \eqref{eq:multi_GEV}.  

Specifically, on the basis of the marginal domain of attraction (see \ref{eq:GEV}) and the approximation in \eqref{eq:GPD_approx} for each marginal distribution, we define the transformations
\begin{equation}\label{eq:marginal_transform}
z_i:=z_i(\cdot)\equiv z_i(\cdot;\btheta) = \frac{k}{n}\left(1+\gamma_i\frac{\cdot-\mu_i}{\sigma_i}\right)_{+}^{-1/\gamma_i},\quad i=1,2,
\end{equation}
where $\btheta=(\btheta_1,\btheta_2)$.
From the bivariate domain of attraction \eqref{eq:multi_GEV} and the max-stability property \citep[Ch~6]{dHF06},
for large $\by$, where $\by=\ba_n\bx+\bb_n$, we have
\begin{equation*}\label{eq:app_biv_doa}
F(\by)\approx G(\by;\bvartheta^*),
\end{equation*}
where $\bvartheta^*=(\btheta{^{*}}, h)$, $\btheta^*=(\btheta_1^*,\btheta_2^*)$, $\btheta_i^*=(\mu_i^*,\sigma_i^*,\gamma_i)$, $i=1,2$, and where $h$ is the angular density. Replicating the arguments involving the Pareto distribution for deriving the approximations in \eqref{eq:gpd} and \eqref{eq:gev_app} for each margin, then for $\by\geq \bt$, where $\bt=(t_1,t_2)$ is a large threshold, we obtain the further approximation
\begin{equation}\label{eq:second_app_biv_doa}
F(\by)\approx \exp\left(-L(\bz;\bvartheta)\right)=:\widetilde{G}(\bz;\bvartheta),\qquad \by\geq\bt,
\end{equation}
where $\bz=(z_1,z_2)$ with $z_i=z_i(y_i;\btheta)$  given by \eqref{eq:marginal_transform} and where
$\bvartheta=(\btheta,h)$.

Following \cite{MPAV2016} we model the angular density using Bernstein polynomials.
Writing $L(\bz)=(z_1+z_2)A(v)$ with
$v=z_2/(z_1+z_2)$, we model Pickands dependence function $A(v)$ through a Bernstein polynomial of degree $\kappa=0,1,\ldots$ as
\begin{equation}\label{eq:poly_pick}
A_\kappa(v;\bbeta_\kappa)=\frac{1}{\kappa+1}\sum_{j=0}^\kappa \beta_j\dbeta(v;j+1,\kappa-j+1),
\end{equation}
where $\bbeta_\kappa=(\beta_1,\ldots,\beta_\kappa)^\top$ is a parameter vector satisfying suitable conditions so that $A_\kappa$ in \eqref{eq:poly_pick} defines a valid Pickands dependence function \citep[][Section 3.1]{MPAV2016}, and  $\dbeta(\cdot;a,b)$ is the beta density function with parameters $a > 0$ and $b > 0$. 
Modelling Pickands dependence function with a polynomial in Bernstein form 
is equivalent to modelling the angular distribution with a Bernstein polynomial. 
The corresponding angular density in Bernstein form is then
\begin{equation}\label{eq:ang_dens}
h_{\kappa-1}(w;\beeta_\kappa)=\sum_{j=0}^{\kappa-2}(\eta_{j+1}-\eta_j)\dbeta(w;j+1,\kappa-j-1),
\end{equation}
where $w\in[0,1]$ and the elements of $\beeta_\kappa=(\eta_0,\ldots,\eta_{\kappa-2})^\top$ must satisfy suitable conditions so that $h_{\kappa-1}$ in \eqref{eq:ang_dens} is a valid angular density \citep[][Section 3.1]{MPAV2016}. The vectors of coefficients $\bbeta_\kappa$  and   $\beeta_\kappa$ are related via a one-to-one relationship \citep[][Proposition 3.2]{MPAV2016}.

Let $(\by_1,\ldots,\by_n)$ be a sample of independent bivariate observations from a distribution 
$F$ on $\R^2_+$, for which $F\in\MDA(G)$. Then, exploiting the approximation $\widetilde{G}$ in \eqref{eq:second_app_biv_doa} we construct the censored likelihood function
$$
\Lik(\bvartheta) = \prod_{i=1}^n\Lik(\by_i;\bvartheta),
$$
where 
each likelihood contribution depends on the domain where $\by_i$ falls. Specifically, 
\begin{equation}
\label{eq:biv_cens_lik}
\Lik(\by;\bvartheta) \propto
\begin{cases}
\widetilde{G}(\bz;\bvartheta)|_{\bz=\bz(\bt)}, & \mbox{ if } \by\leq \bt\\[1em]
- \frac{\partial z_1(y_1)}{\partial y_1} \widetilde{G}(z_1,z_2;\bvartheta) L^{(z_1)}(z_1,z_2;\bvartheta)|_{z_1=z_1(y_1),z_2=z_2(t_2)}, & \mbox{ if } y_1>t_1, y_2\leq t_2\\[1em]
- \frac{\partial z_2(y_2)}{\partial y_2} \widetilde{G}(z_1,z_2;\bvartheta) L^{(z_2)}(z_1,z_2;\bvartheta)|_{z_1=z_1(t_1),z_2=z_2(y_2)}, & \mbox{ if } y_1\leq t_1, y_2> t_2\\[1em]
\mathcal{J}\cdot\widetilde{G}(\by;\bvartheta)(L^{(z_1)}(\bz;\bvartheta)L^{(z_2)}(\bz;\bvartheta)- L^{(z_1,z_2)}(\bz;\bvartheta)|_{\bz=\bz(\by)}, & \mbox{ if } \by>\bt,
\end{cases}
\end{equation}
where
$$
\mathcal{J}=\frac{\partial z_1(y_1)}{\partial y_1}\cdot\frac{\partial z_2(y_2)}{\partial y_2},
$$
\begin{eqnarray*}
L^{(z_1)}(\bz;\bvartheta)=\frac{\partial}{\partial z_1}L(\bz;\bvartheta) &=& A_\kappa(v;\bbeta_\kappa)-vA_\kappa'(v;\bbeta_\kappa),\\[0.5em]
L^{(z_2)}(\bz;\bvartheta)=\frac{\partial}{\partial z_2}L(\bz;\bvartheta) &=& A_\kappa(v;\bbeta_\kappa)+(1-v)A_\kappa'(v;\bbeta_\kappa),\\[0.5em]
L^{(z_1,z_2)}(\bz;\bvartheta)=\frac{\partial^2}{\partial z_1\partial z_2}L(\bz;\bvartheta) &=& -\frac{1}{z_1+z_2}v(1-v)A_\kappa''(v;\bbeta_\kappa),
\end{eqnarray*}
and where $v=z_2/(z_1+z_2)$,  $z_i(\cdot;\btheta)$, $i=1,2$ \eqref{eq:marginal_transform},  $A'_\kappa$ and $A''_\kappa$ are the first and second derivatives of $A_\kappa$  with respect to $v$, and where we re-express the full vector of unknown parameters as $\bvartheta=(\btheta_1,\btheta_2,\kappa,\bbeta_\kappa)$. 
We infer the marginal GEV parameters and the extremal dependence structure through a semiparametric Bayesian approach. In particular, we combine the inferential scheme for each marginal parameter set $\btheta_1, \btheta_2$ described above with the trans-dimensional MCMC  scheme for inferring the dependence structure over the unknown number of elements in $\bbeta_\kappa$ suggested by \citet{MPAV2016} (see also \citealt{a-villalobos+w13,sisson+f11}). 
The latter takes into account that at each MCMC iteration the dimension of $\bbeta_\kappa$ changes with $\kappa$ and that the size of $\bbeta_\kappa$ is potentially infinite \citep[see][Section 3.2 for details]{MPAV2016}.

While it is more convenient to derive the likelihood function using the representation of extremal dependence through Pickands dependence function \citep{MPAV2016}, it is simpler to define a prior distribution on the angular distribution. In this case, the prior distribution for $\bbeta_\kappa$ can be deduced from the prior distribution on $\beeta_\kappa$ by exploiting the one-to-one relationship between the two \citep{MPAV2016}. Accordingly, the MCMC algorithm may be implemented using the likelihood function parametrised through Pickands dependence function and the prior distribution for this model that is induced from the prior on the angular distribution.

Recall that for all $\kappa \geq3$, $\beeta_\kappa$ must satisfy some suitable constraints so that $h_{\kappa-1}$ in \eqref{eq:ang_dens} is a valid angular density. Specifically, the elements of $\beeta_\kappa$ must be a nondecreasing sequence in $[0,1]$ such that their sum is equal to $\kappa/2$ \citep[][Proposition 3.1]{MPAV2016}. Then, following \cite{MPAV2016}, for any fixed $\kappa \geq3$, the prior on  $(\beeta_\kappa,\kappa)$, with $\beeta_\kappa = (\eta_0, \ldots, \eta_{\kappa-1})$, is defined as
\begin{equation}
\label{eqn:etaPrior}
\Pi(\beeta_\kappa, \kappa) = \Pi(\eta_1, \ldots, \eta_{\kappa-2}| p_0, p_1, \kappa) \Pi(p_1 | \kappa, p_0) \Pi(p_0)\Pi(\kappa),
\end{equation}
where $\eta_0 = p_0$ and $\eta_{\kappa-1} = 1-p_1$ represent the point masses at the endpoints of the simplex ($H(\{0\})$ and $H(\{1\})$),
$\Pi(\kappa) = \mathrm{NegBin}(\kappa -3 | m_{NB}, \sigma_{NB})$
with mean $m_{NB}>0$ and variance $\sigma_{NB}>0$,
$\Pi(p_0) = \textrm{Unif}(0, 1/2)$ and $\Pi(p_1 | \kappa, p_0) = \textrm{Unif}(a,b)$ with $a = \max\{ 0, (\kappa-1)p_0 - \kappa/2 + 1\}$ and $b=(p_0-\kappa/2-1)/(\kappa-1)$. The conditional prior distribution on the remaining parameters is set as
$$
\Pi(\eta_1, \ldots, \eta_{\kappa-2}| p_0, p_1, \kappa) 
= \textrm{Unif}(\mathcal{D}_\beeta),
$$
where the domain $\mathcal{D}_\beeta$ is a suitable set such that the elements of $\beeta_\kappa$ satisfy the required conditions above specified \citep[see][Section 3.2 for details]{MPAV2016}. Notice that, although the theory in Section \ref{ssec:biv_quantiles} assumes models for which the condition $H(\{0\})=H(\{1\})=0$ holds, our proposed inferential method is capable of estimating point masses at the vertices $0$ and $1$, if this exists. 
Within the trans-dimensional MCMC update, the pair $(\beeta_{\kappa^{(j)}}^{(j)}, \kappa^{(j)})$ at time $j$ is updated through the proposal distribution 
$$
q(\beeta_\kappa, \kappa | \beeta_{\kappa^{(j)}}^{(j)}, \kappa^{(j)} )
=  q_\beeta(\beeta_\kappa | \kappa)q_\kappa(\kappa | \kappa^{(j)}) 
$$
where
$q_\beeta(\beeta_\kappa | \kappa) = \Pi(\beeta_\kappa | \kappa)$ is the conditional prior implied by \eqref{eqn:etaPrior}, and $q_\kappa(\kappa|\kappa^{(j)})$ is defined such that if $\kappa^{(j)}=3$, it places mass on $\kappa=4$ with probability $1$ and if $\kappa^{(j)}>3$ it places mass on $\kappa^{(j)} - 1$ and $\kappa^{(j)} + 1$ with equal probability. Using $q_\beeta(\beeta_\kappa | \kappa) = \Pi(\beeta_\kappa | \kappa)$ means that these terms cancel in the between-model acceptance probability, whether implemented under the $\beeta_\kappa$ or $\bbeta_\kappa$ parameterisation.

The full MCMC sampler is summarised in Algorithm \ref{alg:algo_joint}, where $\Lik \left(\btheta_1, \btheta_{2}, \kappa, \bbeta_\kappa\right)$ indicates the bivariate censored likelihood \eqref{eq:biv_cens_lik}. 
Separate Robbins-Monro RWMH updates are implemented for each set of marginal parameters $\btheta_i, i=1,2$, and the above scheme for the dependence parameter updates.
See \citet[][Section 3.2 ]{MPAV2016} for further details.

\begin{algorithm}[h!]
\caption{Trans-dimensional MCMC scheme}
\label{alg:algo_joint}
\textbf{Initialize:} Set $\bvartheta^{(0)} = \left(\btheta_1^{(0)}, \btheta_2^{(0)}, \kappa^{(0)}, \bbeta_{\kappa^{(0)}}^{(0)}\right),\beeta_{\kappa^{(0)}}^{(0)}$, $\tau_i^{(0)}$ and $\Sigma_i^{(0)}$ for $i=1,2$. \;
\SetKwRepeat{REPEAT}{repeat}{until}
\For{$j = 0$ \To{} $M$}{
% STEP 1 & 2
% 
\textbf{Step 1:} {\em Marginal component 1:} \;
\quad Draw proposal $\btheta_1' \sim MVN(\btheta_1^{(j)}, \tau^{(j)}_1 \Sigma^{(j)}_1)$.\;
\quad Compute acceptance probability 
$\pi_1 = \min \left(  
 \frac{\Lik \left(\btheta_1', \btheta_{2}^{(j)}, \kappa^{(j)}, \bbeta_{\kappa^{(j)}}^{(j)} \right)\Pi(\btheta_1') }
       	{\Lik \left(\btheta_1^{(j)}, \btheta_{2}^{(j)}, \kappa^{(j)}, \bbeta_{\kappa^{(j)}}^{(j)} \right)\Pi(\btheta_1^{(j)}) }, 1 \right).$\;
       	\quad Draw $U_1 \sim \mathcal{U}(0,1)$. If $\pi_1 > U_1$ then set $\btheta_1^{(j+1)} = \btheta_1'$ else set $\btheta_1^{(j+1)} = \btheta_1^{(j)}$.\;
\quad Update $\Sigma^{(j)}_1$ according to \eqref{eq:A_update}. \;
\quad Update $\tau^{(j)}_1$ according to \eqref{eq:updatetau}. \;
\textbf{Step 2:} {\em Marginal component 2:} \;
\quad Draw proposal $\btheta_2' \sim MVN(\btheta_2^{(j)}, \tau^{(j)}_2\Sigma^{(j)}_2)$.\;
       	\quad Compute acceptance probability 
       	$\pi_2 = \min \left(  
       	\frac{\Lik \left(\btheta_1^{(j+1)}, \btheta_{2}', \kappa^{(j)}, \bbeta_{\kappa^{(j)}}^{(j)} \right)\Pi(\btheta_2') }
       	{\Lik \left(\btheta_1^{(j+1)}, \btheta_{2}^{(j)}, \kappa^{(j)}, \bbeta_{\kappa^{(j)}}^{(j)} \right)\Pi(\btheta_2^{(j)}) } , 1 \right).$\;
       	\quad Draw $U_2 \sim \mathcal{U}(0,1)$. If $\pi_2 > U_2$ then set $\btheta_2^{(j+1)} = \btheta_2'$ else set $\btheta_2^{(j+1)} = \btheta_2^{(j)}$.\;
        \quad Update $\Sigma^{(j)}_2$ according to \eqref{eq:A_update}. \;
        \quad Update $\tau^{(j)}_2$ according to \eqref{eq:updatetau}. \;
       \textbf{Step 3:} {\em Dependence structure:} \;
       \quad Draw proposal $\kappa' \sim q_\kappa(\kappa | \kappa^{(j)})$ and $\beeta_{\kappa'}' \sim q_\eta(\beeta_\kappa | \kappa')$, and compute $\bbeta_{\kappa'}'$.\;
\quad Set $c = 1/2$ if $\kappa^{(j)}=3$ or $c = 1$ if $\kappa^{(j)}>3$.\;
\quad Compute acceptance probability 
$\pi_3 = \min \left(  c
       \frac{\Pi (\kappa')}{\Pi (\kappa^{(j)})}
       \frac{\Lik \left(\btheta_1^{(j+1)}, \btheta_2^{(j+1)}, \kappa', \bbeta_{\kappa'}' \right) }
       {\Lik \left(\btheta_1^{(j+1)}, \btheta_2^{(j+1)}, \kappa^{(j)}, \bbeta_{\kappa^{(j)}}^{(j)} \right) } , 1 \right).$\;
       \quad Draw $U_3 \sim \mathcal{U}(0,1)$. If $\pi_3 > U_3$ then set $\kappa^{(j+1)}=\kappa', \beeta_{\kappa^{(j+1)}}^{(j+1)}=\beeta_{\kappa'}'$, $\bbeta_{\kappa^{(j+1)}}^{(j+1)}=\bbeta_{\kappa'}'$ else set $\kappa^{(j+1)}=\kappa^{(j)}, \beeta_{\kappa^{(j+1)}}^{(j+1)}=\beeta_{\kappa^{(j)}}^{(j)}$, $\bbeta_{\kappa^{(j+1)}}^{(j+1)}=\bbeta_{\kappa^{(j)}}^{(j)}$. \;
       }
\end{algorithm}
%

%

% =====================================================
% =====================================================
\section{Simulation experiments}
\label{sec:simu}
% =====================================================
% =====================================================

% =====================================================
% =====================================================
\subsection{Univariate}
\label{ssec:univ_sim}
% =====================================================
% =====================================================

We generate $n=1500$ observations from each of three distributions: $\mbox{Fr\'{e}chet}$ with location, scale and shape parameters equals to $\psi_0=3$, $\varsigma_0=1$ and $\xi_0=1/3$, respectively; $\mbox{Half-}t$ with scale and degrees of freedom equal to $\sigma_0=1$ and
$\nu_0=1/3$; $\mbox{Inverse Gamma}$ with shape parameters equal to $\eta_0=1/2$ and
$\lambda_0=1$. 
We recall that $\text{Fr}(y;\psi,\varsigma,\xi)=\exp(-((y-\psi)/\varsigma)^{-\xi})$, with $y,\xi>0$, is the Fr\'{e}chet family of distributions with scale $\varsigma>0$ and location $\psi<y$ parameters. 
These distributions are in the domain of attraction of a GEV distribution with tail indices 
$\gamma_0 = 3, 3$ and $2$, respectively  \citep[][Ch.~2]{BGST04}. The univariate likelihood function \eqref{eq:uni_CensLik} is used, censoring observations below the $90$-th empirical quantile. We specify a prior distribution for $\btheta$ as a product of 
uniform prior distributions on the real line for $\mu$, $\log(\sigma)$ and $\gamma$, i.e.~$\Pi(\btheta)\equiv\Pi(\mu,\sigma,\gamma):=\Pi(\mu)\Pi(\log(\sigma))\Pi(\gamma)$. This improper prior distribution, with
$\Pi(\mu,\sigma,\gamma)\propto 1/\sigma$ with $\sigma>0$, 
leads to a proper posterior distribution \citep{northrop2016}. We run the MCMC sampler for $M=50,000$ iterations for each dataset. 

\begin{figure}[t!]
%
% FRECHET
%
\includegraphics[width=0.24\textwidth, page=1]{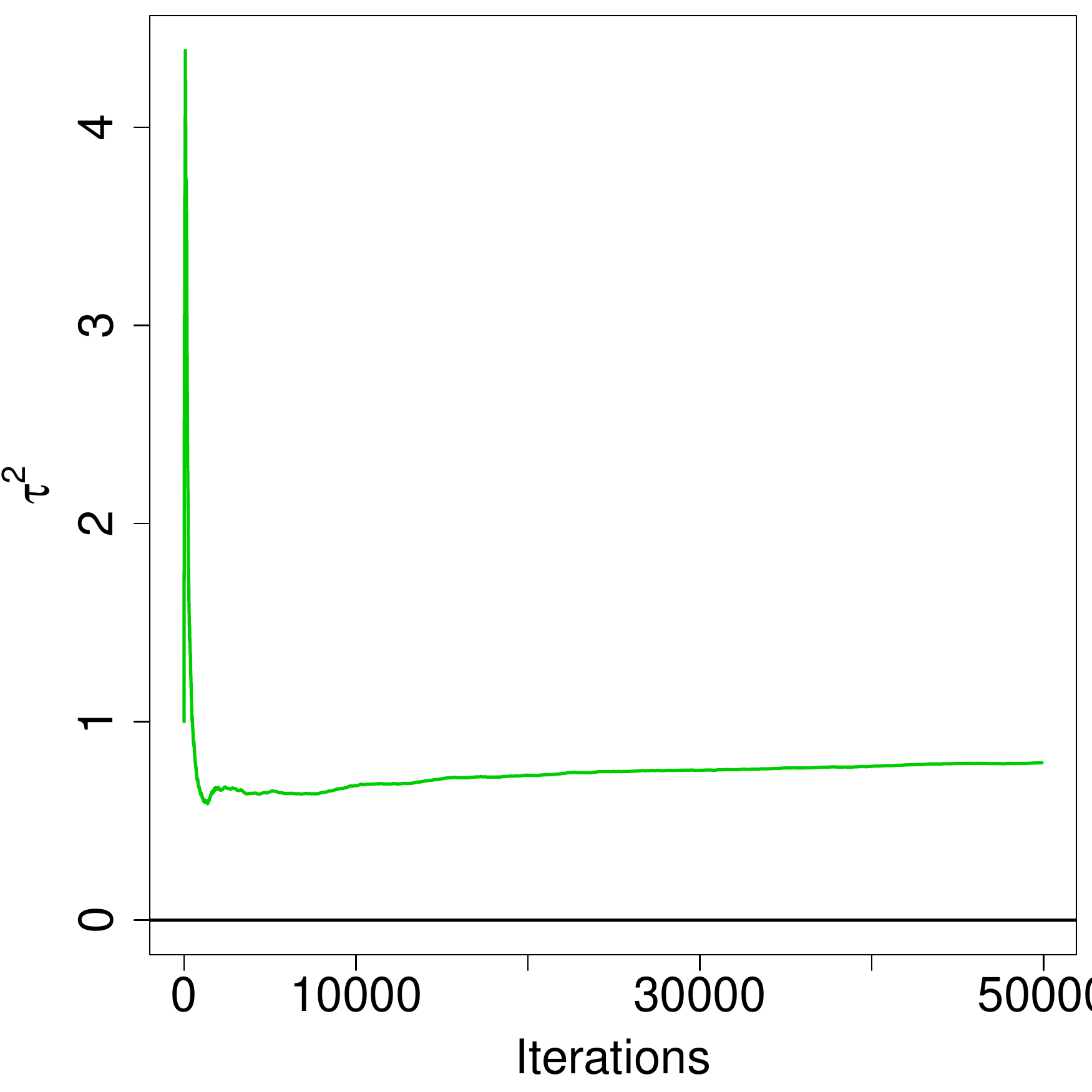}
\includegraphics[width=0.24\textwidth, page=2]{FRE_n1500_prob09_loc30_scale10_shape3_3_diagnostics_new.pdf}
\includegraphics[width=0.24\textwidth, page=3]{FRE_n1500_prob09_loc30_scale10_shape3_3_diagnostics_new.pdf}
\includegraphics[width=0.24\textwidth, page=4]{FRE_n1500_prob09_loc30_scale10_shape3_3_diagnostics_new.pdf} \\
%
% HALF-T
%
\includegraphics[width=0.24\textwidth, page=1]{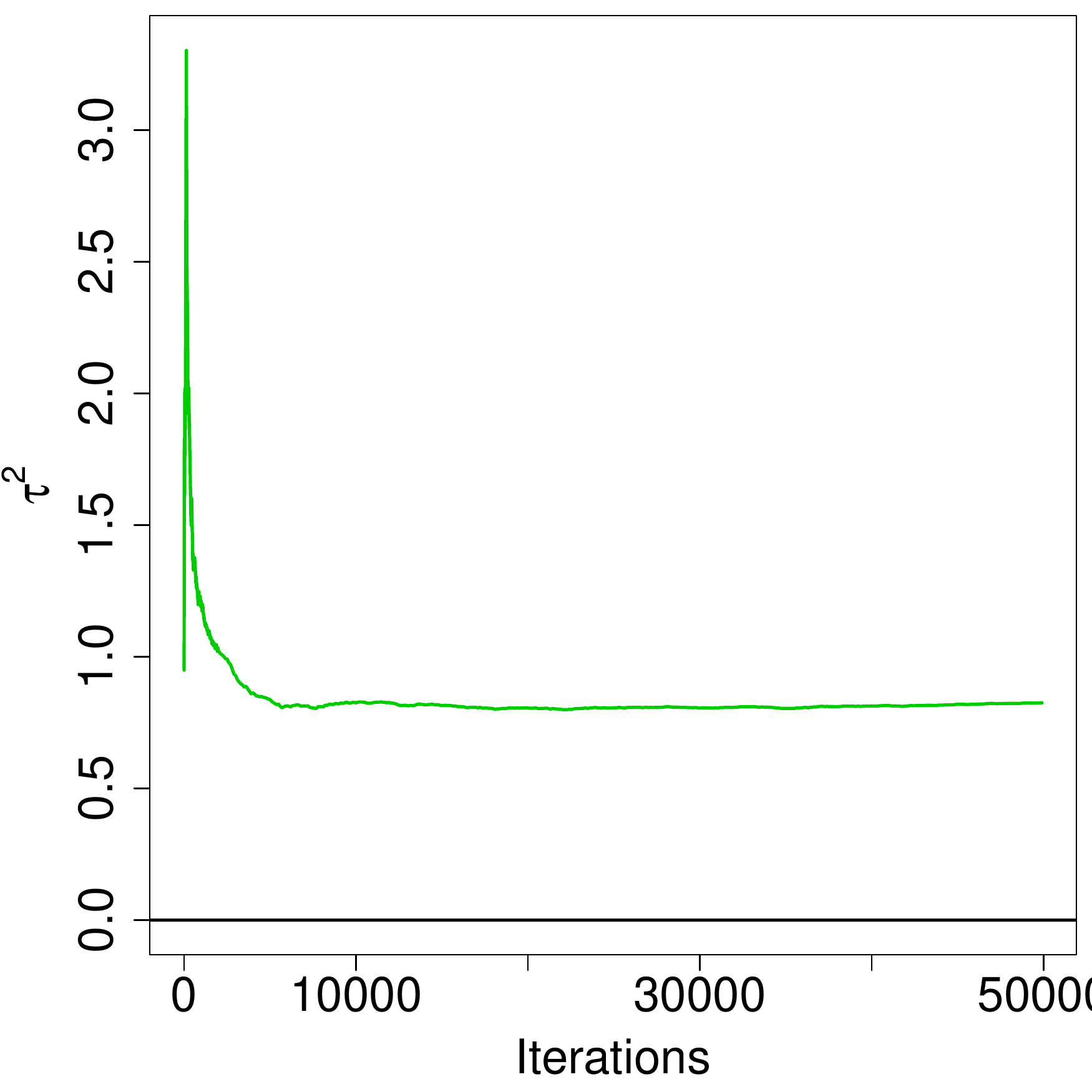}
\includegraphics[width=0.24\textwidth, page=2]{HT_n1500_prob09_Dof3_33333333333333_sigma1_diagnostics_new.pdf}
\includegraphics[width=0.24\textwidth, page=3]{HT_n1500_prob09_Dof3_33333333333333_sigma1_diagnostics_new.pdf}
\includegraphics[width=0.24\textwidth, page=4]{HT_n1500_prob09_Dof3_33333333333333_sigma1_diagnostics_new.pdf} \\
%
% INVERSE-GAMMA
%
\includegraphics[width=0.24\textwidth, page=1]{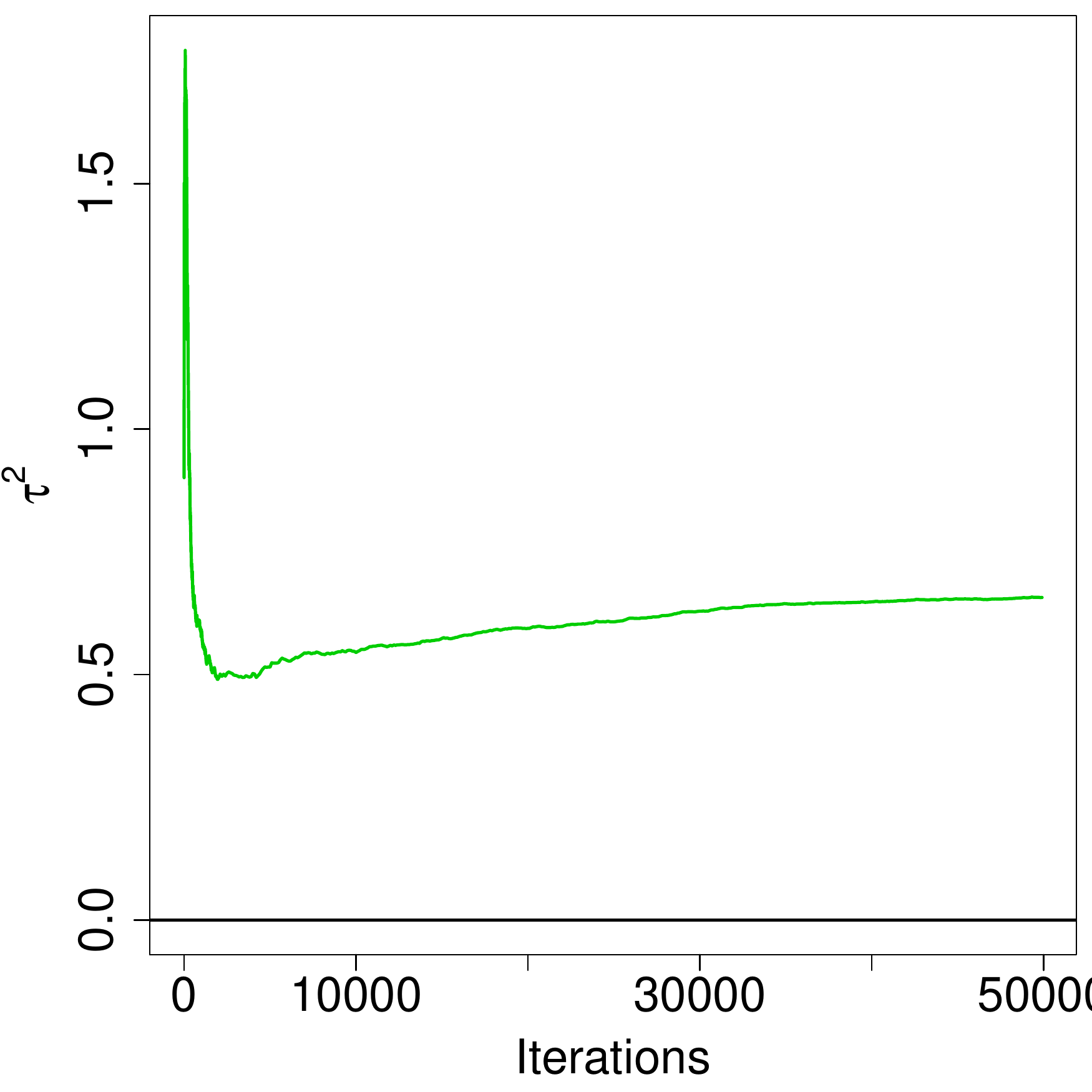}
\includegraphics[width=0.24\textwidth, page=2]{IG_n1500_prob09_scale10_shape5_diagnostics_new.pdf}
\includegraphics[width=0.24\textwidth, page=3]{IG_n1500_prob09_scale10_shape5_diagnostics_new.pdf}
\includegraphics[width=0.24\textwidth, page=4]{IG_n1500_prob09_scale10_shape5_diagnostics_new.pdf} \\
\caption{\small Univariate extreme quantile region results for $n=1,500$ Fr\'{e}chet (top row), Half-$t$ (middle) and inverse-Gamma (bottom) distributed data. Columns illustrate sampler scaling parameter $\tau^2$ (left) and overall acceptance probability against sampler iteration (centre-left) with target sampler acceptance rate of $\pi^*=0.234$ indicated by horizontal line.
Centre-right column shows histogram and kernel density estimates of tail index $\gamma$  after removing $m=30,000$ iterations burn-in. Crosses are the lower and upper bounds of the estimated $95\%$ credibility interval.
Right column illustrates log-scale posterior densities of quantiles corresponding to the exceedance probabilities $p=1/750$ (light grey), $1/1500$ and $1/3000$ (dark grey).
Posterior mean and true tail index are indicated by dashed and solid lines respectively, observed data indicated by points on the $x$-axis.
}
\label{fig:univ}
\end{figure}

Each row in Figure~\ref{fig:univ} shows the estimation results for each dataset: Fr\'{e}chet (top), Half-$t$ (middle) and inverse Gamma (bottom).
The columns on the left  present trace plots of the scaling parameter $\tau^2$ (initialised at $\tau^{(0)}=1$) and the  sampler average acceptance probability. Through the Robbins-Monro process both quantities converge rapidly, with the sampler acceptance rate effectively achieving the target (``optimal'') acceptance rate of $\pi^*=0.234$ (solid horizontal line) after no more than $m=30,000$ iterations, which we remove as sampler burn-in.
The centre-right panels illustrate histogram and kernel density estimates of the posterior distribution of the tail index $\gamma$, with dashed and solid vertical lines indicating the posterior mean and true value, respectively. The crosses along the horizontal axis show the lower and upper bounds of the estimated $95\%$ credibility interval.
In each case the posterior for $\gamma$ puts most of its mass where the true value lies.

Finally, the rightmost panels display the estimated posterior densities of quantiles corresponding to the exceedance probabilities $p=1/750$ (light grey), $1/1500$ and $1/3000$ (dark grey), derived from \eqref{eq:extreme_q_approx}. Vertical dashed and solid lines again represent the estimated posterior mean and true quantile values with the latter that are on the log scale: $19.86$, $21.94$, $24.02$ for the Fr\'{e}chet; $18.73$, $20.81$, $22.89$ for the Half-$t$; $13.48$, $14.87$, $16.25$ for the Inverse Gamma. The corresponding estimated central $95\%$ credible intervals are: $(17.67, 22.73)$, $(19.41, 25.36)$, $(21.16, 27.95)$; $(17.22, 22.38)$, $(18.93, 25.00)$, $(20.68, 27.60)$ and  $(12.58, 16.58)$, $(13.86, 18.58)$, $(15.14, 20.57)$, respectively. 
The points on the $x$-axis are the upper $5\%$ of the observed dataset. As the sample size is $n=1500$, the largest value is a realisation of an event occurring with probability $1/1500$, corresponding to the second investigated quantile. 
For example, on the log scale, the largest observation from the Fr\'{e}chet distribution (top row of Figure~\ref{fig:univ}) is $24.36$.
This means that the probability of observing an event taking the value $24.36$ or greater is less than $1/750$, and is more likely to be an event with probability closer to $1/3000$.
In all cases, the true tail indices and quantiles are always included inside the estimated central $95\%$ credible intervals, confirming the accuracy of our proposed method for estimating such quantities.
In Section 2 of the Supplementary Material we investigate the sensitivity of the above results to variations in the experimental setup. In particular we consider a higher censoring threshold (at the 95-th empirical quantile) and alternative prior specifications for $\mu, \sigma, \gamma$. 
Performance is qualitatively similar to the above in each case.
%

% =====================================================
% =====================================================
\subsection{Bivariate}
\label{ssec:biv_sim}
% =====================================================
% =====================================================

Estimating and quantifying bivariate extreme quantile regions is more challenging than the univariate setting.
Here we examine data simulated from two distributions defined on $\R^2_+$: 
the positive bivariate Cauchy
and a positive truncated bivariate-$t$ density. The former has been previously considered in the literature \citep[e.g.][]{EdHK2013,CEdH11}, and we consider the second one as an alternative flexible model. Two further examples (the so-called Asymmetric and Clover densities) are investigated in the Supplementary Material (Section 3).
As for the univariate setting, we simulate $n=1500$ observations from each density and marginally censor all observations that fall below the corresponding $90$-th marginal empirical quantile. We are again interested in estimating quantile regions for events with probability $p=1/750$, $1/1500$ and $1/3000$, corresponding to regions that contain very little or no observed data.
Details about the selected distributions are:
\begin{itemize}
\item {\bf Bivariate Cauchy distribution on $\R_+^2$:} The Cauchy probability density function is
$$
f(\bx) = \frac{2}{\pi \left( 1 + x_1^2 + x_2^2 \right)^{3/2}}, \quad \bx \in \R_+^2.
$$
The Cauchy distribution is very heavy-tailed with tail indices $\gamma_1 = \gamma_2 =1$, and angular density $h(w) = 2^{-1} \left( w^2 + (1-w)^2 \right)^{-3/2}$. The angular basic density function is $q_*(w) = \left( w^2 + (1-w)^2 \right)^{1/2}$ and the associated basic set  is
$$
\cS = \left\{ \bx\in\R^2_+: r > \left( w^2 + (1-w)^2 \right)^{-1/2}, \, w \in [0, 1] \right\}.
$$

\item {\bf Truncated bivariate-$t$ distribution on $\R^2_+$:} We consider a truncated two-dimensional Student-$t$ 
distribution with unit-variances, correlation $0\leq\rho\leq 1$ and $\nu>0$ degrees of freedom. 
The bivariate-$t$ probability density function is
$$
f(\bx) = \frac{t_{2, \nu}\left( \bx; \rho \right)}{T_{2, \nu}\left( \bzero; \rho \right)}, \quad \bx \in \R_+^2,
$$
where $t_{d, \nu}$ denotes the $d$-dimensional centred Student-$t$ probability density function with correlation $0\leq \rho\leq 1$ and $\nu>0$ degrees of freedom, $T_{2, \nu}$ is the $d$-dimensional centred Student-$t$ distribution function, and $\bzero=(0,0)^\top$. The tail indices are  $\gamma_1 = \gamma_2 = 1/\nu$ and 
vary with the degree of freedom and the angular density is
$$
h(w) = \frac{1}{2 \nu w^3} \sqrt{\frac{\nu +1}{1-\rho^2}} \left( \frac{1-w}{w}\right)^{\frac{1-\nu}{\nu}} \frac{ t_{1,\nu+1}\left(  \sqrt{\frac{\nu +1}{1-\rho^2}} \left(  \left( \frac{1-w}{w}\right)^{\frac{1}{\nu}} - \rho \right) \right)}{1- T_{1,\nu+1} \left( - \rho \sqrt{\frac{\nu +1}{1-\rho^2}} \right) } .
$$
The angular basic density is equal to
$$
q_*(w) = \left( \nu w^{-(1+2/\nu)} \sqrt{\frac{\nu +1}{1-\rho^2}} 
\frac{ t_{1,\nu+1} \left(  \sqrt{\frac{\nu +1}{1-\rho^2}} \left(  \left( \frac{1-w}{w}\right)^{\frac{1}{\nu}} - \rho \right) \right)}
{1- T_{1,\nu+1} \left( - \rho \sqrt{\frac{\nu +1}{1-\rho^2}} \right)}
\right)^{-\frac{\nu}{\nu + 2}},
$$
and the associated basic set is
$$
\cS = \left\{ \bx\in\R^2_+: r > \left( \frac{\nu}{w^{1+\frac{2}{\nu}}} 
\sqrt{\frac{\nu +1}{1-\rho^2}} 
\frac{ t_{1,\nu+1} \left(  \sqrt{\frac{\nu +1}{1-\rho^2}} \left( \frac{(1-w)^{\frac{1}{\nu}}}{w^{\frac{1}{\nu}}} - \rho \right) \right)}
{1- T_{1,\nu+1} \left( - \rho \sqrt{\frac{\nu +1}{1-\rho^2}} \right)}
\right)^{\frac{\nu}{\nu +2}}, w \in [0, 1] \right\}.
$$
Details concerning the derivation of the angular density are available in Appendix~\ref{app:t_pos_quad}. 
It is well known that a bivariate-$t$ distribution defined on $\R^2$ is in the domain of attraction of
the  Extremal-$t$ model \citep[see e.g.][]{beranger2015}. The angular distribution corresponding to this model places mass on all the subsets of the unit simplex, which in the bivariate case corresponds to the interval $(0,1)$
and the vertices $\{0\}$ and $\{1\}$. However, when considering a truncated bivariate-$t$ distribution defined on $\R^2_+$, it can be shown that the corresponding angular distribution does not put mass at the endpoints, i.e.~$H(\{0\}) = H(\{1\}) = 0$ (see Appendix~\ref{app:t_pos_quad}).
Note that when $\nu=1$ and $\rho=0$ the truncated bivariate-$t$ distribution reduces to the bivariate Cauchy on $\R^2_+$. 
\end{itemize}

\begin{figure}[t]
\includegraphics[width=\textwidth]{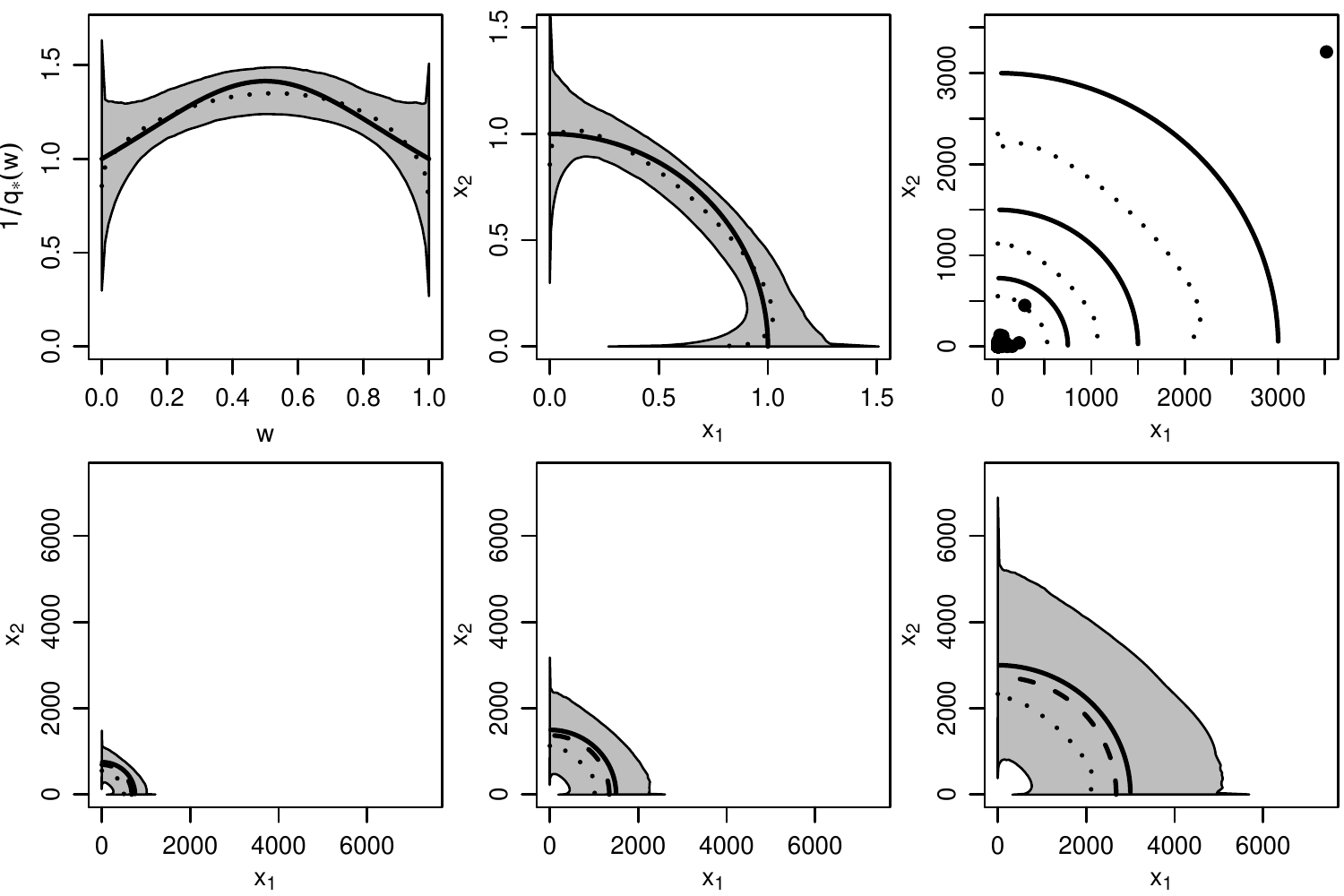}
\caption{\small 
Extreme quantile analysis for the Cauchy distribution. True (solid lines), posterior mean estimate (dotted) and $90\%$ credible regions (grey) for the inverse of the angular basic density (top left), the basic set $\cS$ (top middle), and the quantiles with probability $p=1/750, 1/1500$ and $1/3000$ (bottom left to bottom right). Dashed lines in bottom panels shows the EdHK point estimator. Top right panel illustrates the simulated dataset (points) and true and posterior mean estimated quantiles.}
\label{fig:Cauchy}
\end{figure}

\begin{figure}[t]
\includegraphics[width=\textwidth]{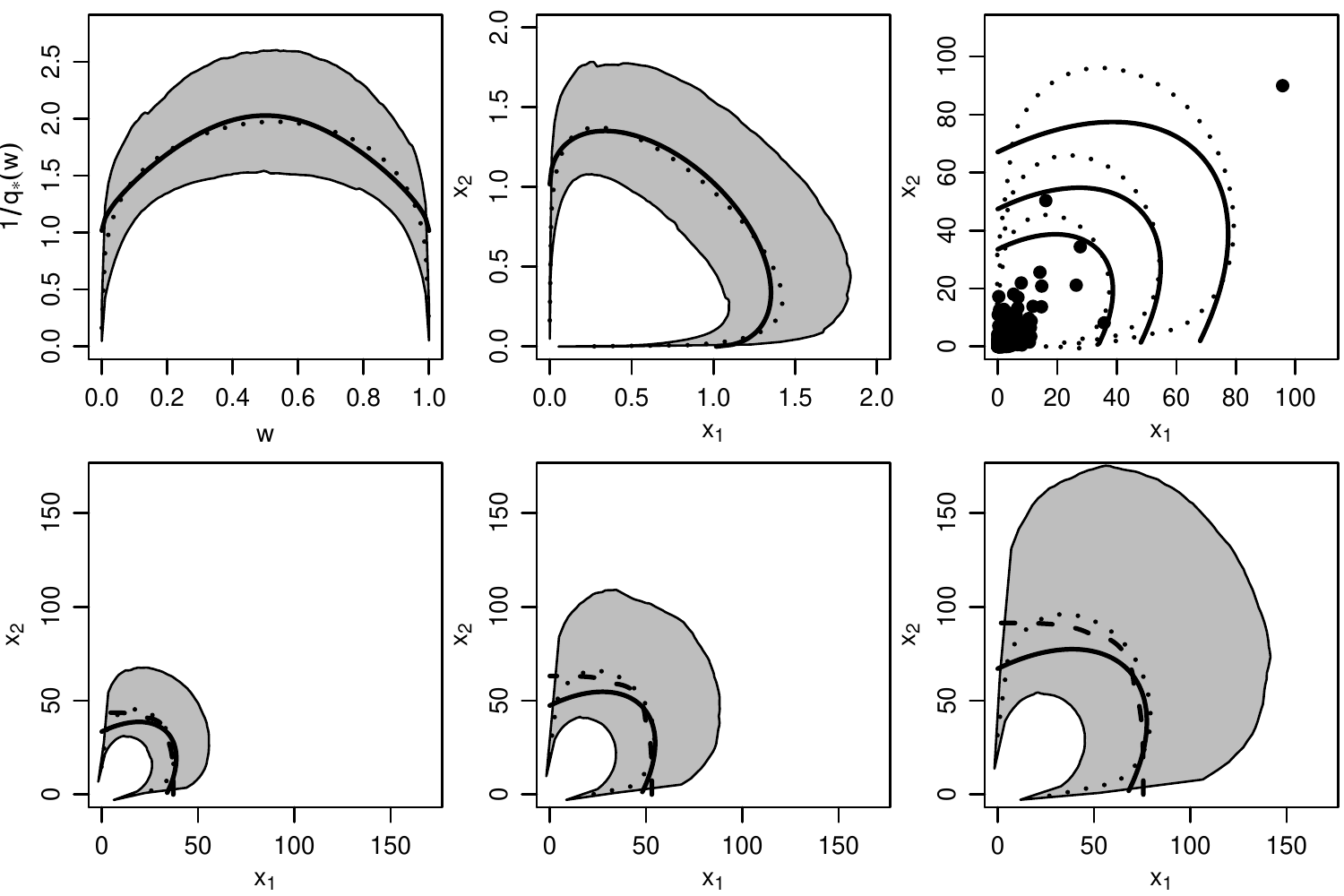}
\caption{\small 
Extreme quantile analysis for the bivariate-$t$ distribution with $\nu_0=2$ degrees of freedom and correlation $\rho_0=0.5$. True (solid lines), posterior mean estimate (dotted) and $90\%$ credible regions (grey) for the inverse of the angular basic density (top left), the basic set $\cS$ (top middle), and the quantiles with probability $p=1/750, 1/1500$ and $1/3000$ (bottom left to bottom right). Dashed lines in bottom panels shows the EdHK point estimator. Top right panel illustrates the simulated dataset (points) and true and posterior mean estimated quantiles.
}
\label{fig:HT}
\end{figure}

As before, we specify a uniform prior distribution $\Pi(\mu_i,\sigma_i,\gamma_i)$ for each margin $i=1,2$.
As the dependence structures for the above models are known to be smooth (see solid lines in the top left panels of Figures~\ref{fig:Cauchy}--\ref{fig:HT}),
it is expected they can be well modelled through relatively low degree polynomials. Hence we set the prior distribution for the polynomial degree as $\Pi(\kappa)=\mathrm{NegBin}(m_{NB}=3.2, \sigma_{NB}=4.48)$. 
Even though the selected models do not permit mass in the corners of the simplex (i.e.~$H(\{0\}) = H(\{1\}) = 0$), in the analysis
we still allow for the possibility of non-zero point masses at the endpoints of the simplex by specifying $\Pi(p_0) = \Pi(p_1) = \textrm{Unif}(0, 0.1)$. This prior is slightly different to those in \cite{MPAV2016} to better represent our prior belief that $p_0$ and $p_1$ are likely to be small for these data. See \cite{MPAV2016} for an analysis of alternative prior specifications.

We run Algorithm \ref{alg:algo_joint} for $M=50,000$ iterations and determine the burn-in period $m$ by visual inspection of trace plots of the marginal scaling parameters $\tau_1, \tau_2$ and the overall acceptance rates of the marginal ($\pi_i^*=0.234, i=1,2$) and dependence proposals. 

For each draw from the posterior distribution, we can construct the angular density $h_{\kappa-1}(w;\beeta)$ using \eqref{eq:ang_dens}, combine this with the marginal shape indices $\gamma_1, \gamma_2$ into $q(w)$ \eqref{eq:dens_exp_measure}, and compute the angular basic density $q_*(w)$
(see Section \ref{ssec:biv_quantiles} for details). 
Hence, for each fixed $w \in (0,1)$, we obtain samples from the posterior of the angular basic density $q_*(w)$. Its inverse $q_*^{-1}(w)$ and pointwise central credible intervals are shown in the top-left panels of Figures~\ref{fig:Cauchy}--\ref{fig:HT}.

The boundary of the basic set $\cS$ corresponds to those points $(x,y)\in(0,\infty)^2$ such that $x+y = q_*^{-1}(w)$, i.e.~the points $\left(w q_*^{-1}(w), (1-w) q_*^{-1}(w) \right)$. For fixed $w\in(0,1)$, the posterior distribution of this boundary is available, from which we may calculate the posterior mean and 90\% central credible intervals. Computed over all $w\in(0,1)$ the (pointwise) posterior mean and credible intervals for the basic set are illustrated in the top-centre panels of Figures~\ref{fig:Cauchy}--\ref{fig:HT}.

To construct the extreme quantile regions for a given small probability $p$ (top-right panel and bottom row of Figures~\ref{fig:Cauchy}--\ref{fig:HT}),
 consider a point $w \in (0,1)$. 
As before, we may construct the posterior distribution of the points $(x,y)=\left(w q_*^{-1}(w), (1-w) q_*^{-1}(w) \right)\in\R^2_+$
with angle $w$ and radial component $q_*^{-1}(w)$, as an estimate of a point on the boundary of $\cS$. 
For each $(x,y)$ point in this posterior, we may compute the quantile $\widetilde{Q}_n$ associated with probability $p$ via  \eqref{eq:Q_n_approx}, leading to a posterior distribution approximating the $1-p$ bivariate quantile level at a particular (estimated) point in $\cS$.
We then use 
the marginal $0.05$ and $0.95$ quantiles of this bivariate posterior to define an approximate $90\%$ credible region, and the marginal posterior means to produce a mean estimate. This procedure is repeated for other $w \in (0,1)$. 

The top right panel in Figures~\ref{fig:Cauchy}--\ref{fig:HT} provides a comparison between the true extreme quantile region (solid line) and the posterior mean estimate (dotted line), for $p=1/750, 1/1500$ and $1/3000$, with the observed data (points) overlaid. The bottom panels in each figure illustrate each extreme quantile region separately, but with the estimated 90\% pointwise credible regions and the point estimates of the quantile regions given in \citet{EdHK2013} (dotted lines) are included for comparison. We refer to the latter as the EdHK estimator.

Figure~\ref{fig:Cauchy} illustrates the results  for the Bivariate Cauchy distribution.
The estimated (pointwise posterior mean) inverse angular basic density  $q_*^{-1}(w)$ (dotted line; top-right panel) describes the behaviour of the true function (solid line) well, and is fully included in the (pointwise) estimted $90\%$ credible regions. 
The curvature in the dependence structure around $w=0.5$ is not as pronounced as the true $q_*^{-1}(w)$, which  consequently impacts on the estimated basic set and extreme quantile regions. The bottom panels show that the mean extreme quantile levels (dotted) line are close to the true levels (solid line) and provide a similar fit to the EdHK point estimate (dashed) for each probability $p=1/750, 1/1500$ and $1/3000$. In addition all three curves consistently appear in the centre of the $90\%$ credible regions. 

Finally, Figure~\ref{fig:HT} presents analysis for the truncated bivariate-$t$ distribution.
The dependence structure $q_*^{-1}(w)$ is very accurately estimated within the interior of the simplex, although
towards the endpoints  the estimated $q_*^{-1}(w)$ appears to approach $0$ whereas the true value is approximately $1$.  Producing quantile regions that seem drawn to the origin for low values of either component, may at first appear erroneous. However, inspection of the simulated dataset (top-right panel) reveals that there are no extreme observations that lie along the axes, thereby explaining the behaviour of our estimator. 
On balance, the estimated posterior means follow the general shape of the true quantile regions and provide similar results than  the competitor EdH estimates (dashed lines). The true quantile regions
are amply included in the (pointwise) estimated $90\%$ credible regions.

Overall, the proposed methodology is able to accurately estimate both marginal and  dependence structure of process extremes.
As a point estimator, the posterior mean of the extreme quantile region performs well in recovering the true region, while being responsive to the dependence structure within the data itself, as with the bivariate-$t$ data analysis (Figure \ref{fig:HT}).
It also seems to perform more consistently than  the EdHK estimator (e.g.~Figures~5 and 6 of the Supplementary Material). 
The credible regions both provide some measure of the uncertainty inherent with low probability events, while also 
allowing for judgements regarding whether
exceptionally high observations can still be considered to belong to events with particular probabilities (e.g.~whether the single large outlier in Figure~\ref{fig:HT} can be considered a $1/1500$ or $1/3000$ probability event).

% =====================================================
% =====================================================
\section{Analysis of extreme air pollution levels in Milan}
\label{sec:data_analysis}
% =====================================================
% =====================================================

Understanding the extremes of air pollutants is of critical importance, especially in the context of climate change \citep{desario2013}. In recent work, \citet{martins2017} used standard univariate extreme value theory to compare the air quality between the two largest urban regions in Brazil, \citet{heffernan2004} and \citet{beranger2015} estimated the extremal dependence between pollutants in Leeds, U.K., \citet{coles1996} studied the extreme temporal behaviour of airborne NO$_2$ particles in Milan between 1984--1994, and \citet{vettori2018} performed a spatial analysis of air pollution  in Los Angeles, U.S.A.

Here we study extreme air pollution levels  recorded in Milan, Italy, over the winter period October 31st--February 27/28th, between December 31st 2001 and December 30th 2017. We examine the daily mean level of PM$_{10}$ and daily maximum levels of NO, NO$_2$ and SO$_2$. These data were also considered by \citet[][Table~3]{falk2018} although there the objective was estimating joint exceedance probabilities (with excesses in all  variables).

\begin{figure}[t!]
% FRECHET
\includegraphics[width=0.24\textwidth, page=2]{Pollutants_vs_MaxTemp.pdf}
\includegraphics[width=0.24\textwidth, page=1]{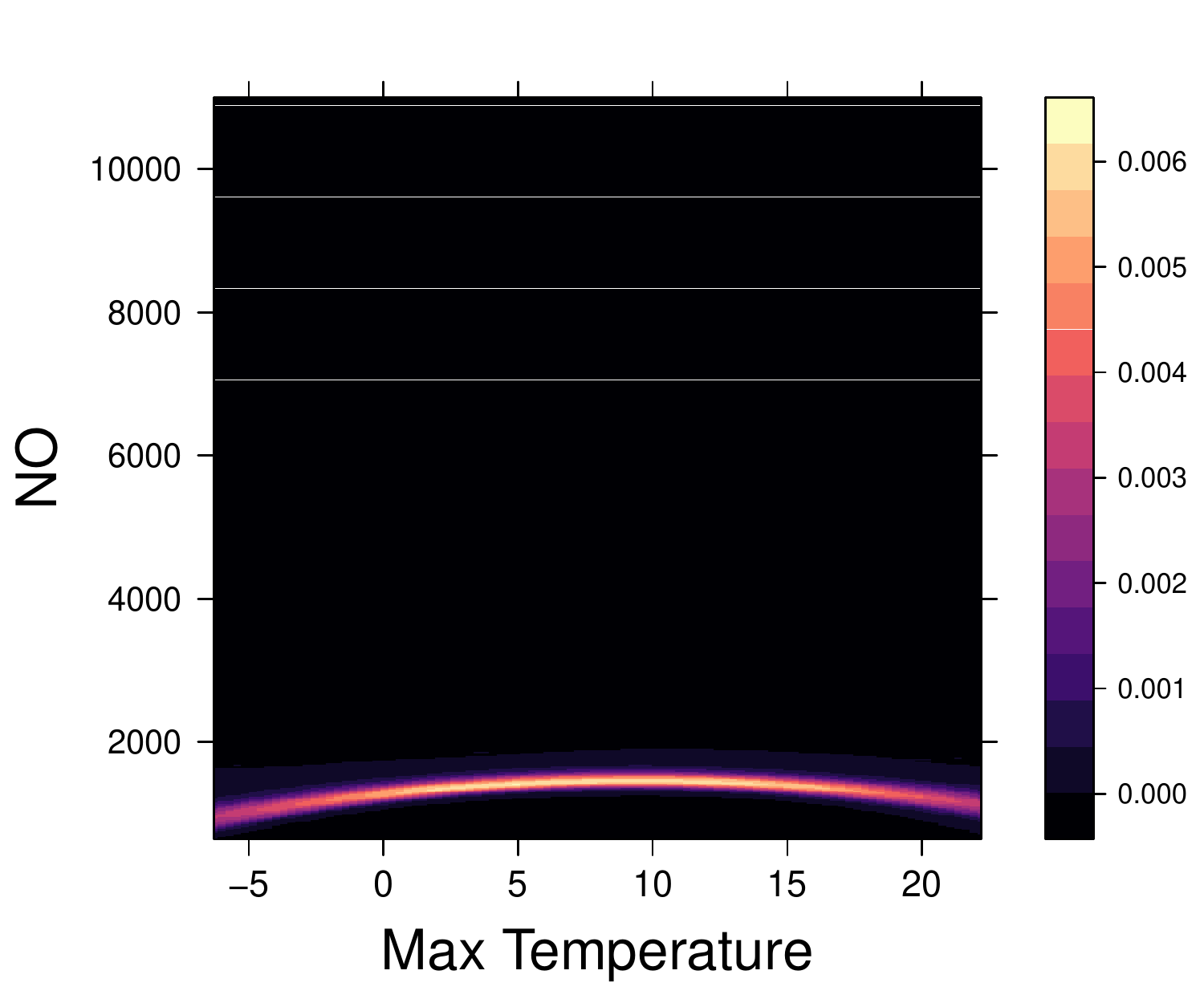}
\includegraphics[width=0.24\textwidth, page=2]{QuantilePlot_winter_NO_covMaxTemp_prob09_new.pdf}
\includegraphics[width=0.24\textwidth, page=3]{QuantilePlot_winter_NO_covMaxTemp_prob09_new.pdf} \\
\includegraphics[width=0.24\textwidth, page=1]{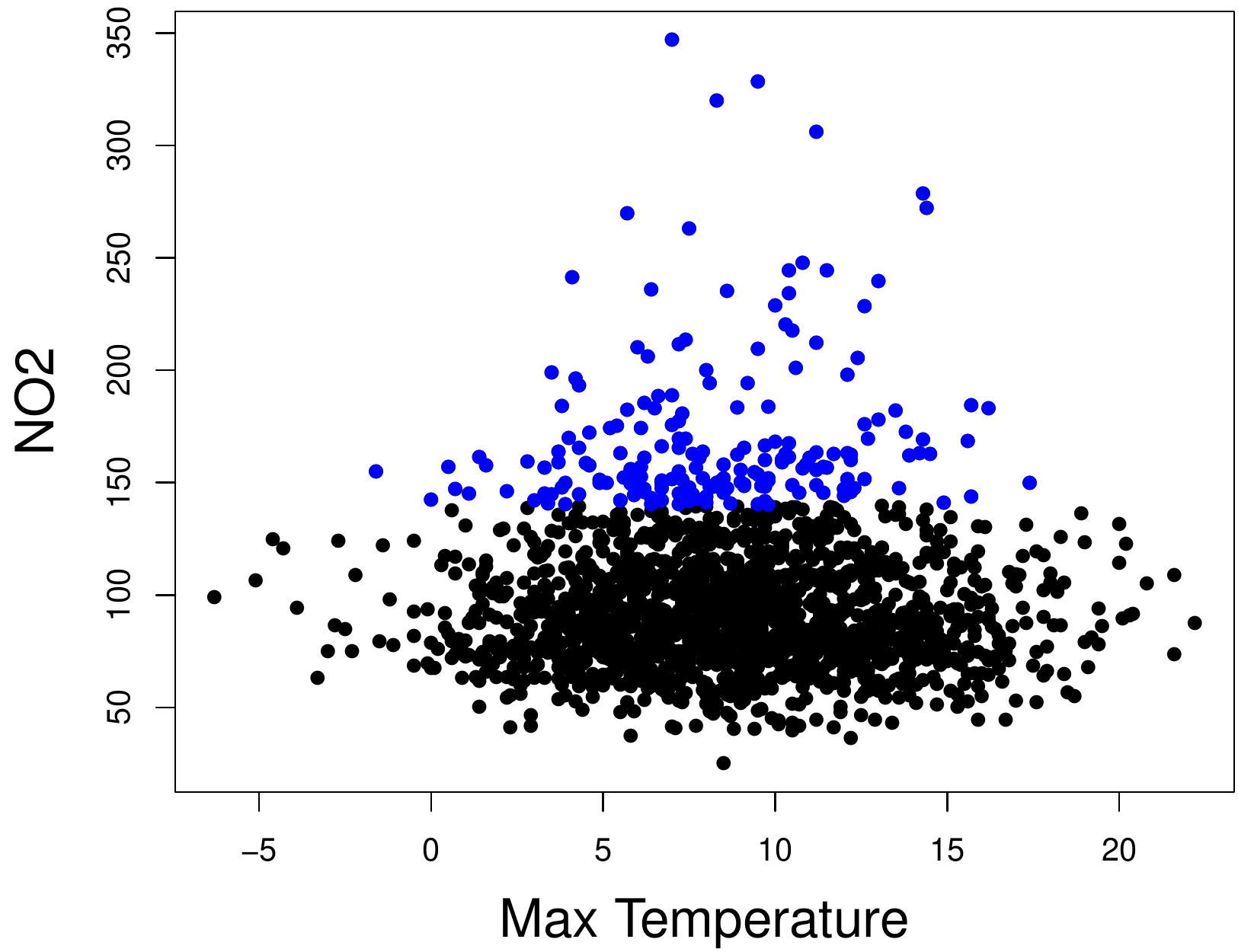}
\includegraphics[width=0.24\textwidth, page=1]{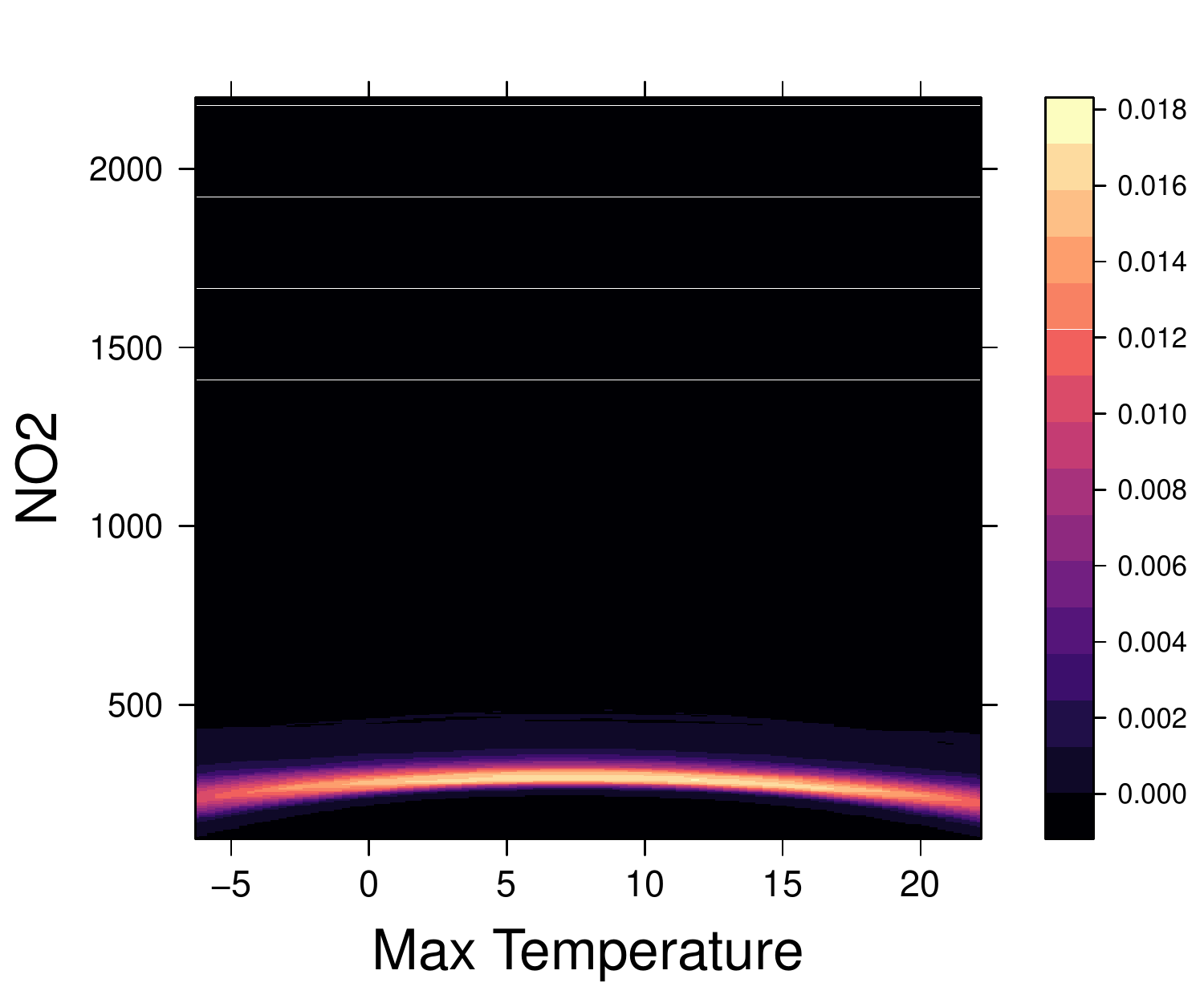}
\includegraphics[width=0.24\textwidth, page=2]{QuantilePlot_winter_NO2_covMaxTemp_prob09_new.pdf}
\includegraphics[width=0.24\textwidth, page=3]{QuantilePlot_winter_NO2_covMaxTemp_prob09_new.pdf} \\
\includegraphics[width=0.24\textwidth, page=4]{Pollutants_vs_MaxTemp.pdf}
\includegraphics[width=0.24\textwidth, page=1]{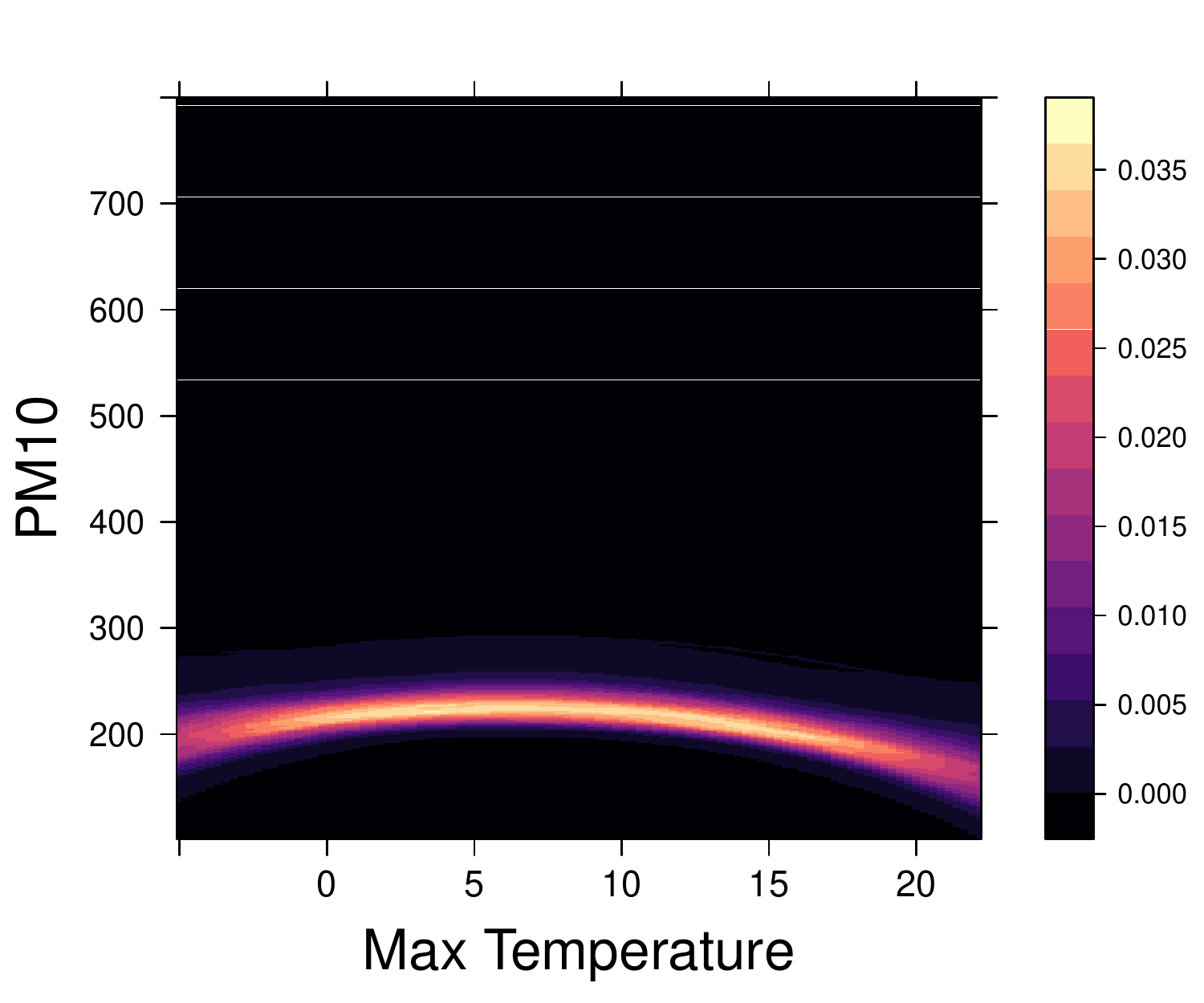}
\includegraphics[width=0.24\textwidth, page=2]{QuantilePlot_winter_PM10_covMaxTemp_prob09_new.pdf}
\includegraphics[width=0.24\textwidth, page=3]{QuantilePlot_winter_PM10_covMaxTemp_prob09_new.pdf} \\
\includegraphics[width=0.24\textwidth, page=3]{Pollutants_vs_MaxTemp.pdf}
\includegraphics[width=0.24\textwidth, page=1]{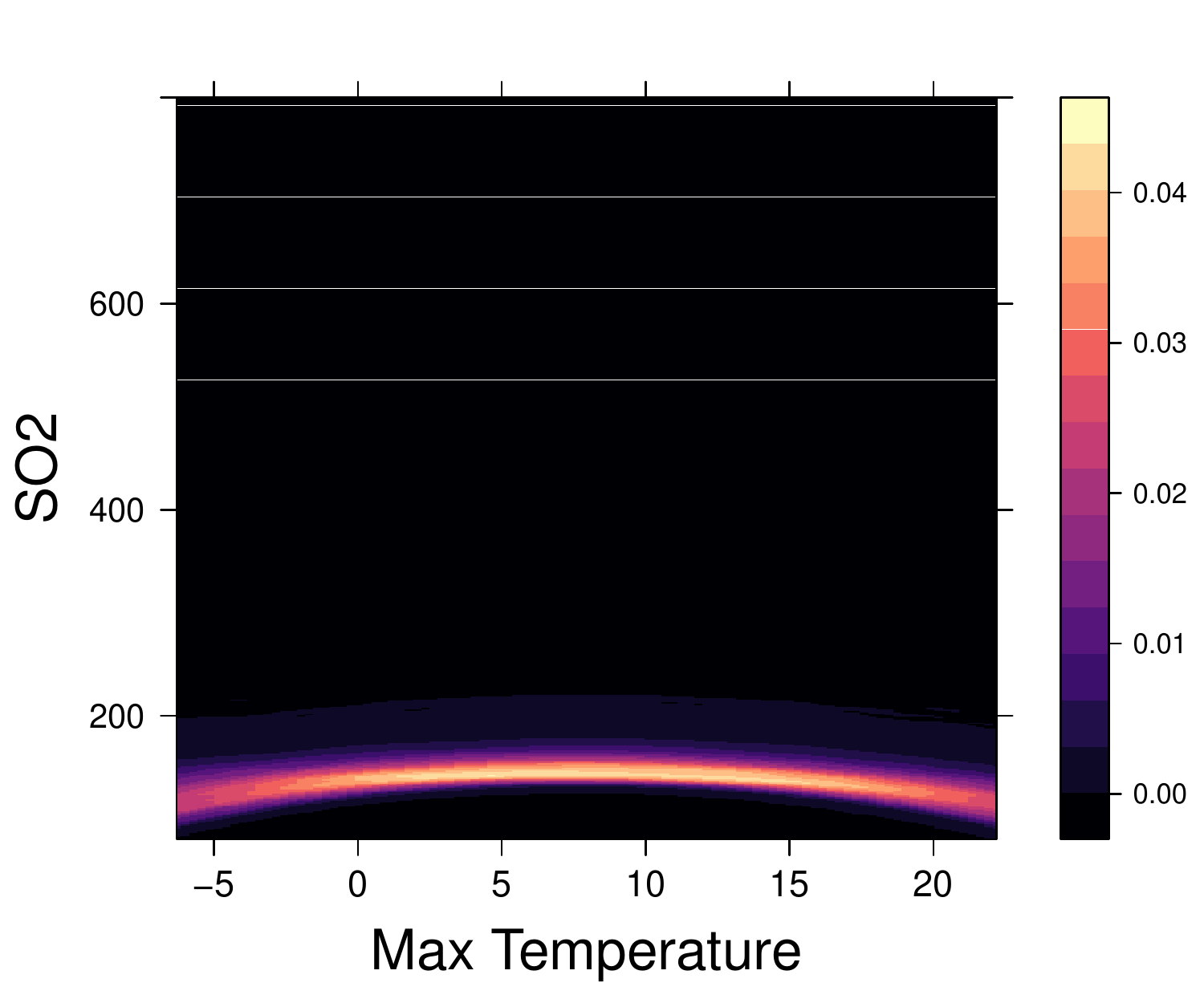}
\includegraphics[width=0.24\textwidth, page=2]{QuantilePlot_winter_SO2_covMaxTemp_prob09_new.pdf}
\includegraphics[width=0.24\textwidth, page=3]{QuantilePlot_winter_SO2_covMaxTemp_prob09_new.pdf}
\caption{\small Univariate analyses of (top row to bottom row) NO, NO$_2$, PM$_{10}$ and SO$_2$ data
relative to the maximum daily temperatures ($x$-axes).
Left column: Scatterplots of each pollutant versus maximum daily temperature, with observations above the threshold shown in blue. Second to fourth columns: Image plots of the univariate distribution of quantiles associated with probability $p=1/600, 1/1200$ and $1/2400$ as a function of maximum daily temperature.}
\label{fig:univ_data}
\end{figure}

Here, the aim is to estimate bivariate extreme quantile regions  given by 
a high concentration of the two pollutants and with a small probability of occurrence. 
This analysis is important for understanding the long term impact of air quality on health. When modelling univariate extremes it is common to express marginal parameters via a regression model which may be functions of space or other covariates \citep{padoan+rs10}. 
For air pollution in particular, evidence is overwhelming that interactions between the pollution and temperatures cannot be disregarded \citep{desario2013,cheng2012,katsouyanni1993, roberts2004}.
The leftmost panels of Figure~\ref{fig:univ_data} show scatterplots of each pollutant against the daily maximum temperature. 
All four indicate a possible quadratic relationship between pollutant and maximum temperature, and so we write the $i$-th marginal mean $\mu_i = \beta_{0,i} + \beta_{1,i} z + \beta_{2,i} z^2$ as a quadratic function of the maximum temperature $z$. 
Other covariates could have been included if they were available, and regressions on $\sigma_i$ and $\gamma_i$ constructed, although we did not pursue that here.
For observations that fall below the threshold (black points in left panels of Figure~\ref{fig:univ_data}),
as these observation are censored it could be argued that the level of the covariates at the threshold should be used when evaluating the likelihood contribution. This would beneficially reduce computational costs. However, as the covariates are still available for censored observations, they still provide valuable information to estimate the regression coefficients. 

Using mean residual life plots \citep[e.g.][Section 1.2.2]{BGST04} we set the marginal $90$-th empirical quantile as the threshold (points above the threshold are blue in Figure~\ref{fig:univ_data}), which results in $t=642.9$ ($n=1796$) for NO, $t=139.5$ ($n=1799$) for NO$_2$, $t=108$ ($n=1779$) for PM$_{10}$ and $t=45.2$ ($n=1809$) for SO$_2$. These thresholds are comparable with those in \citet[][Table~3]{falk2018}.
We specify all marginal prior distributions $\Pi(\mu_i,\sigma_i,\gamma_i)\propto 1/\sigma_i$, for $i=1,\ldots,4$ as in Section \ref{ssec:univ_sim}, and dependence parameter prior distributions as in Section \ref{ssec:infer_biv} with $\Pi(\kappa)=\mbox{NegBin}(\kappa=3|m_{NB}=6,\sigma_{NB}=8)$ to allow for higher degree polynomial modelling of $h_{\kappa-1}(w;\beeta)$ if required. 
For the univariate analysis we implement an MCMC sampler with 300k iterations and retain the final 50k iterations for inference; for the multivariate analysis we retain the final 20k iterations from a chain of length 50k.
The univariate sampler ran in under 1 hour on a single-core  2.2 GHz Intel Core i7 processor on a MacBook Pro; the multivariate sampler took around 2.5 hours.

The image plots in Figure~\ref{fig:univ_data} illustrate the (univariate) posterior distribution of the quantiles associated with the probabilities $p=1/600, 1/1200$ and $1/2400$ (left to right) as a function of the maximum daily temperature, which correspond to pollutant levels that would be expected to be reached once every 5, 10 and 20 winters.
Restricting our interpretation to the range of observed temperature, we note that extreme PM$_{10}$ quantiles (third row) appear higher for lower temperatures rather than high. 
Indeed, the main sources of PM$_{10}$ pollution include combustion engines (both diesel and petrol) and combustion for energy production in households. Accordingly, when maximum temperatures are low, one may expect household heating systems to work at higher capacity and an increase in the use of cars rather than less energy consumptive transport methods such as walking or cycling.

Extreme quantile behaviour for NO$_2$ (second row) as function of maximum temperature appears similar to PM$_{10}$ although the larger range of the quantiles ($y$-axis) reduces the visual curvature.
This behaviour can be explained by the fact that nitrogen dioxide, mainly emitted by power generation, industrial and traffic sources, 
 is an important constituent of particulate matter. Sulfur dioxide (fourth row) forms either naturally (decomposition or combustion of organic matter) or due to human activity (smelting sulfur-containing mineral ores). Accordingly, it is not surprising to observe decreasing extreme SO$_2$ quantiles across temperature levels (fourth row) as maximum temperature moves from $0$ to $20^\circ\mathrm{C}$. The decreasing quantile behaviour as temperature gets cold might be due to a reduction in human activity. 
The posterior mean of the tail indices $\gamma_{\textrm{NO}}, \gamma_{\textrm{NO$_2$}}, \gamma_{\textrm{PM$_{10}$}}$ and $\gamma_{\textrm{SO$_2$}}$ and their corresponding $90\%$ credible interval are $0.050$ $(0.003, 0.138)$, $0.186$ $(0.055,0.337)$, $0.054$ $(0.004, 0.148)$ and $0.064$ $(0.005, 0.167)$ respectively.
 
\begin{figure}[t!]
\centering
\includegraphics[width=0.85\textwidth, page=1]{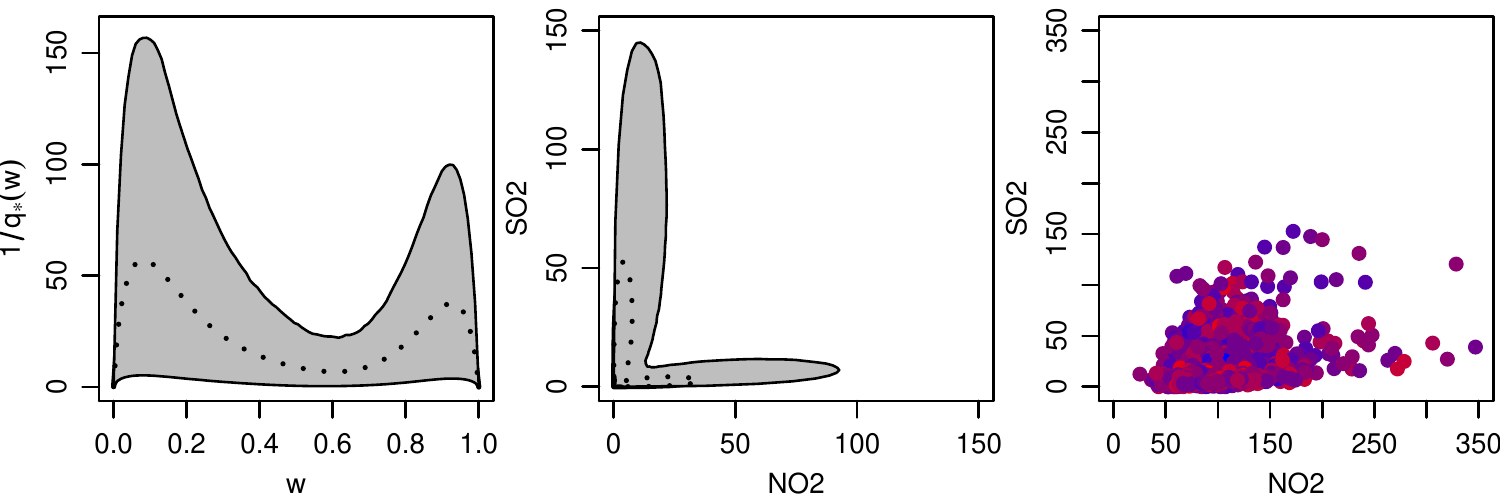} \\
\includegraphics[width=0.85\textwidth, page=2]{Milan_Winter_NO2_SO2_600_1200_2400_dep_new.pdf} \\
\includegraphics[width=0.85\textwidth, page=3]{Milan_Winter_NO2_SO2_600_1200_2400_dep_new.pdf} 
\caption{\small Posterior mean estimate (dotted line) and $90\%$ credible regions (grey) for the inverse of the angular basic density (left column), the basic set $\cS$ (middle column), and and observed data (right column) with temperature dependent data colouring (from cold = blue to warm = red). Top to bottom: The pairs of pollutants shown are NO$_2$/SO$_2$, NO$_2$/PM$_{10}$ and SO$_2$/PM$_{10}$. }
\label{fig:biv_data_dep}
\end{figure}

To date, the link between extreme levels of multiple pollutants and human health has not been well explored, although
a multi-pollutant approach to assess the health risks associated with air pollution has been emerging \citep[e.g.][]{dominici2010,wesson2010}. 
In the following we restrict our attention to the three pollutants with the largest tail indices: NO$_2$, SO$_2$ and PM$_{10}$.

Figure~\ref{fig:biv_data_dep} shows information regarding extremal dependence between the three pollutant pairs: NO$_2$/SO$_2$, NO$_2$/PM$_{10}$ and SO$_2$/PM$_{10}$. The estimated basic sets demonstrate weak dependence between the pairs NO$_2$/SO$_2$ and SO$_2$/PM$_{10}$ and stronger dependence between NO$_2$ and PM$_{10}$. Further, from the pairwise scatterplots it is apparent that neither the coldest (blue dots) nor the warmest (red dots) daily maximum temperatures appear to induce the largest pollutant levels. As discussed above, NO$_2$ is a key constituent of PM$_{10}$ and so it is  realistic to expect the observed strong dependence between these pollutants. Similarly, the sources of PM$_{10}$ (combustion engines) and SO$_2$ (natural or smelting mineral ores) differ, explaining the independence between the extremes of these pollutants to some extent. 
\begin{figure}[t!]
\centering
\includegraphics[width=0.85\textwidth, page=1]{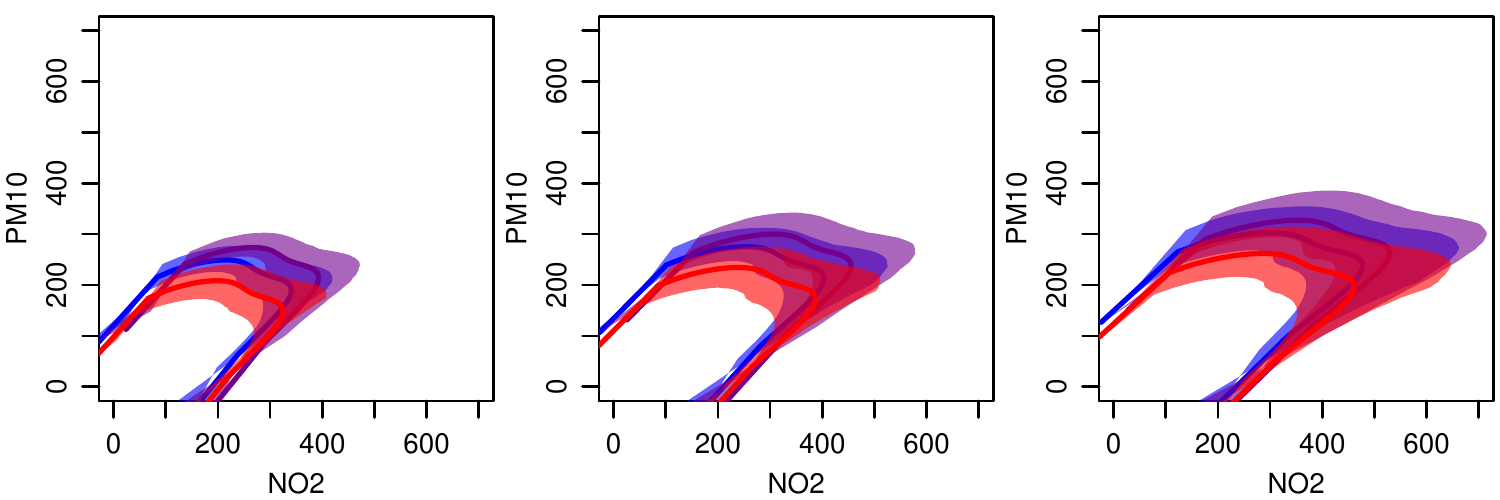}\\
\includegraphics[width=0.85\textwidth, page=2]{Milan_Winter_NO2_PM10_600_1200_2400_quant_new.pdf}\\
\includegraphics[width=0.85\textwidth, page=3]{Milan_Winter_NO2_PM10_600_1200_2400_quant_new.pdf}
\caption{\small Posterior mean estimates (solid line) and $90\%$ credible regions for the extreme bivariate quantiles associated with the probabilities $p=1/600, 1/1200$ and $1/2400$ (left to right) for three maximum daily temperature levels: minimum temperature = blue, median temperature = purple, maximum temperature = red. }
\label{fig:biv_data_quant}
\end{figure}

Figure~\ref{fig:biv_data_quant} illustrates extreme quantiles regions corresponding to events that would expect to occur once every 5, 10 and 20 winters (left to right panels) when the daily maximum temperature is fixed to the minimum, median and maximum observed daily maximum temperatures (blue to red shading). These values are $(-6.3^\circ\textrm{C}, 8.6^\circ\textrm{C}, 22.2^\circ\textrm{C})$, $(-5.1^\circ\textrm{C}, 8.6^\circ\textrm{C}, 22.2^\circ\textrm{C})$ and $(-5.1^\circ\textrm{C}, 8.7^\circ\textrm{C}, 22.2^\circ\textrm{C})$ for the pairs NO$_2$/SO$_2$, NO$_2$/PM$_{10}$ and SO$_2$/PM$_{10}$ respectively. 
The top panels of Figure~\ref{fig:biv_data_quant} exhibit a small quadratic influence of maximum daily temperature on the joint levels of NO$_2$ and SO$_2$, and similarly for SO$_2$ and PM$_{10}$ (bottom row). 
It is apparent that extreme levels of SO$_2$ are reduced with warmer weather. This phenomena can be loosely understood by the fact that sulfur dioxide is an aerosol which cools the planet by reflecting some of the sun's energy back into space. As such, large levels of SO$_2$ are more likely to be associated with cold temperatures. 
Overall the estimated extreme quantile regions capture the behaviour of the data well, with the NO$_2$/PM$_{10}$ pair exhibiting the strongest level of dependence. As expected, as the probability $p$ of an event decreases the quantile regions become larger and reach higher pollutant levels. Despite having observations for 17 consecutive winters, our method is able to provide quantile levels and credible regions for events with probability of occurrence $p$ at arbitrarily low levels. 

In terms of the pollutant thresholds suggested by European emission regulation, 
the top-left panel of Figure \ref{fig:biv_data_quant} shows that
when the the daily maximum temperature is fixed to the maximum, 
we expect that NO$_2$ individually exceeds approximatively $200$ $\mu$g/m$^3$ (i.e.~the short-term limit threshold) once every $5$ winters ($p=1/600$) on average. However, this event is as equally probable as the joint event that both NO$_2$ and PM$_{10}$ each exceed approximatively $200$ $\mu$g/m$^3$. Such a concentration  for PM$_{10}$ is four times higher than the individual limit threshold for this pollutant (see Section \ref{sec:Intro}).
Similarly, for the bottom-left panel of Figure \ref{fig:biv_data_quant},
when the daily maximum temperature is fixed to the median temperature, we expect that PM$_{10}$ individually exceeds approximatively $200$ $\mu$g/m$^3$ on average once every $5$ winters. However,  this event is as equally probable as the joint event that both SO$_2$ and PM$_{10}$ each exceed approximatively $200$ $\mu$g/m$^3$. This concentration for SO$_2$ is almost twice the individual limit threshold for this pollutant (see Section \ref{sec:Intro}). 
These results indicate both the strong dependence between harmful pollutants at extreme levels in Milan,
and that limit threshold alerts for poor air quality in this city are likely to be issued for multiple pollutants simultaneously.

% =====================================================
% =====================================================
\section{Discussion}\label{sec:discussion}
% =====================================================
% =====================================================

We have presented a new method for estimating extreme quantile regions that is responsive to varying levels of extremal dependence, and comes with natural measures of uncertainty, both for model parameters and extreme quantile regions, under the Bayesian paradigm. The method was able to outperform the existing (EdHK) approach of \cite{EdHK2013} which does not provide any measure of uncertainty.
This methodology provides a useful general tool for long-term analysis of the air pollution at the extreme level. We explored this in Section \ref{sec:data_analysis} for assessing and quantifying the health risks associated with multiple extreme pollutant exposures \citep{dominici2010,wesson2010}.

The methods developed in this paper have been incorporated into accessible functions in the {\tt R} Package {\tt ExtremalDep}. This package is freely available from the CRAN repository, see {\tt https://CRAN.R-project.org/package=ExtremalDep}. 
The {\tt R} code for the simulations and the multivariate analysis of real air pollution data 
is  available online at the address {\tt https://www.borisberanger.com/zip/BPS\_2019.zip}.

% =====================================================
% =====================================================
\section*{Acknowledgements}
% =====================================================
% =====================================================

The authors are grateful to Andrea Krajina for sharing the code for the frequentist estimation of bivariate extreme quantile regions and her valuable suggestions and help. SAP is supported by the Bocconi Institute for Data Science and Analytics (BIDSA), Italy and PRIN 2015 research grant {\it Modern Bayesian Nonparametric Methods}. SAS and BB are supported by the Australian Centre of Excellence for Mathematical and Statistical Frontiers (ACEMS; CE140100049) and the Australian Research Council Discovery Project scheme (FT170100079). The authors
are also indebted to the Associate Editor and two anonymous reviewers for their careful reading of the manuscript
and their constructive remarks

%
%%%%%%%%%%%%%%%%%%%%%%%%%%%%%%%%%%%%%%%%%%%%%
%
% Section: APPENDIX
%
%%%%%%%%%%%%%%%%%%%%%%%%%%%%%%%%%%%%%%%%%%%%%
%
\appendix
\section{Appendix}
\subsection{The Extremal-$t$ model with restriction to the positive reals}
\label{app:t_pos_quad}

Consider a student-$t$ distribution restricted to $(0,\infty)$. Using \citet[p.59]{BGST04} the norming constants required in \eqref{eq:GEV} are 
$$
\begin{array}{ccc}
a_n = n^{1/\nu} \left( \frac{ 2 \Gamma\left( \frac{\nu + 1}{2}\right) \nu^{\frac{\nu}{2}-1} }{\sqrt{\pi} \Gamma\left( \frac{\nu}{2}\right) } \right)^{1/\nu}
& \text{and} & b_n =0.
\end{array}
$$
Applying the conditional tail dependence function framework of \citet{nikoloulopoulos2009}, the exponent function can be written as:
\begin{align*}
V(x,y) =& \lim_{n \to \infty} x^{-\nu} \P \left( Z_2 \leq a_n y | Z_1 \leq a_n x \right)
+ y^{-\nu} \P \left( Z_1 \leq a_n x | Z_2 \leq a_n y \right),
\end{align*}
where $(Z_1, Z_2)^\top$ follows a centred bivariate-$t$ distribution on $(0, \infty)^2$ with unit variance, correlation $\rho$ and degree of freedom $\nu$. The conditional distribution of $Z_2|Z_1=z_1$ is a truncated $t$ distribution on $(0, \infty)$ with mean $\rho z_1$, variance $(\nu + z_1^2) (1-\rho^2) / (\nu +1)$ and $\nu + 1$ degrees of freedom. As a consequence, we obtain
$$
\P \left( Z_2 \leq a_n y | Z_1 \leq a_n x \right) = 
\frac{T_{1,\nu+1} \left( \sqrt{\frac{\nu +1}{1-\rho^2}} \frac{a_n (y-\rho x)}{\sqrt{\nu + a_n^2 x^2}} \right) - T_{1,\nu+1} \left( - \rho \sqrt{\frac{\nu +1}{1-\rho^2}} \frac{a_n x}{\sqrt{\nu + a_n^2 x^2}} \right)}
{1- T_{1,\nu+1} \left( - \rho \sqrt{\frac{\nu +1}{1-\rho^2}} \frac{a_n x}{\sqrt{\nu + a_n^2 x^2}} \right) },
$$
and 
$$
\lim_{n \to \infty} \P \left( Z_2 \leq a_n y | Z_1 \leq a_n x \right) =
\frac{T_{1,\nu+1} \left( \sqrt{\frac{\nu +1}{1-\rho^2}} \left( \frac{y}{x}-\rho \right) \right) - T_{1,\nu+1} \left( - \rho \sqrt{\frac{\nu +1}{1-\rho^2}} \right)}
{1- T_{1,\nu+1} \left( - \rho \sqrt{\frac{\nu +1}{1-\rho^2}} \right) }.
$$
Identical calculations can be applied to the second term in the exponent function. Transforming the margins to unit-Fr\'{e}chet margins allows expression of the exponent function as
\begin{align*}
V(x,y) = \frac{1}{1- T_{1,\nu+1} \left( - \rho \sqrt{\frac{\nu +1}{1-\rho^2}} \right) } 
& \left\{ 
\frac{1}{x} \left[ T_{1,\nu+1} \left( \sqrt{\frac{\nu +1}{1-\rho^2}} \left( \frac{y}{x}-\rho \right) \right) - T_{1,\nu+1} \left( - \rho \sqrt{\frac{\nu +1}{1-\rho^2}} \right) \right] \right.\\
& \left. + \frac{1}{y} \left[ T_{1,\nu+1} \left( \sqrt{\frac{\nu +1}{1-\rho^2}} \left( \frac{x}{y}-\rho \right) \right) - T_{1,\nu+1} \left( - \rho \sqrt{\frac{\nu +1}{1-\rho^2}} \right) \right]
\right\}.
\end{align*}
It is easy to verify that $\lim_{y \to 0} \partial / \partial x V(x,y) = 0 $ and $\lim_{x \to 0} \partial / \partial y V(x,y) = 0 $ which implies $H(\{0\}) = H(\{1\}) = 0$. Finally, note that due to the form of $V(x,y)$, taking the double partial derivative with respect to $x$ and $y$ is equivalent to the double partial derivative of the exponent function of the Extremal-$t$ model multiplied by a scaling term. Hence the angular density on the interior of the $2$-dimensional unit simplex is as given.

\bibliographystyle{chicago}
\bibliography{biblio1}

\newpage

\section{Supplementary Material for ``Estimation and uncertainty\\ quantification for 
extreme quantile regions''}

\begin{abstract}
This document contains additional simulation results for evaluating the performance of the Bayesian procedure for inferring univariate and bivariate quantiles.
\end{abstract}

% =====================================================
% =====================================================
\subsection{Introduction}
\label{sec:Intro}
% =====================================================
% =====================================================

Note that all references below of the form ($\cdot^*$) refer to equation ($\cdot$) in the main paper. We recall that ``EdHK'' is the abbreviation for Einmahl-de Haan-Krajina with reference to the results developed in \citet{EdHK2013} 
%

% =====================================================
% =====================================================
\subsection{Complement to Section~4.1}
% =====================================================
% =====================================================

\begin{figure}[t!]
% FRECHET
\includegraphics[width=0.24\textwidth, page=1]{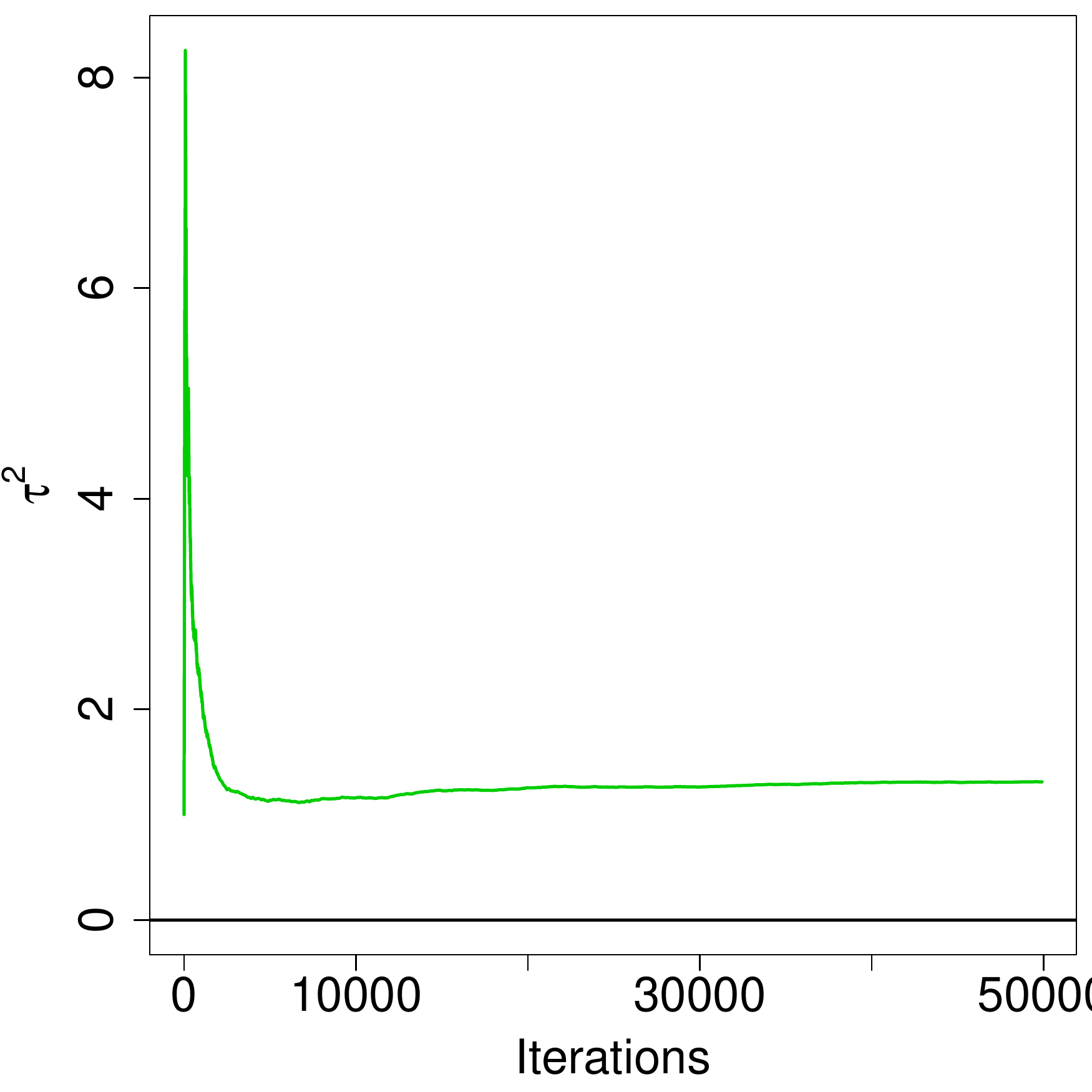}
\includegraphics[width=0.24\textwidth, page=2]{FRE_n1500_prob095_loc30_scale10_shape3_3_diagnostics_new.pdf}
\includegraphics[width=0.24\textwidth, page=3]{FRE_n1500_prob095_loc30_scale10_shape3_3_diagnostics_new.pdf}
\includegraphics[width=0.24\textwidth, page=4]{FRE_n1500_prob095_loc30_scale10_shape3_3_diagnostics_new.pdf} \\
%%
% HALF-T
%
\includegraphics[width=0.24\textwidth, page=1]{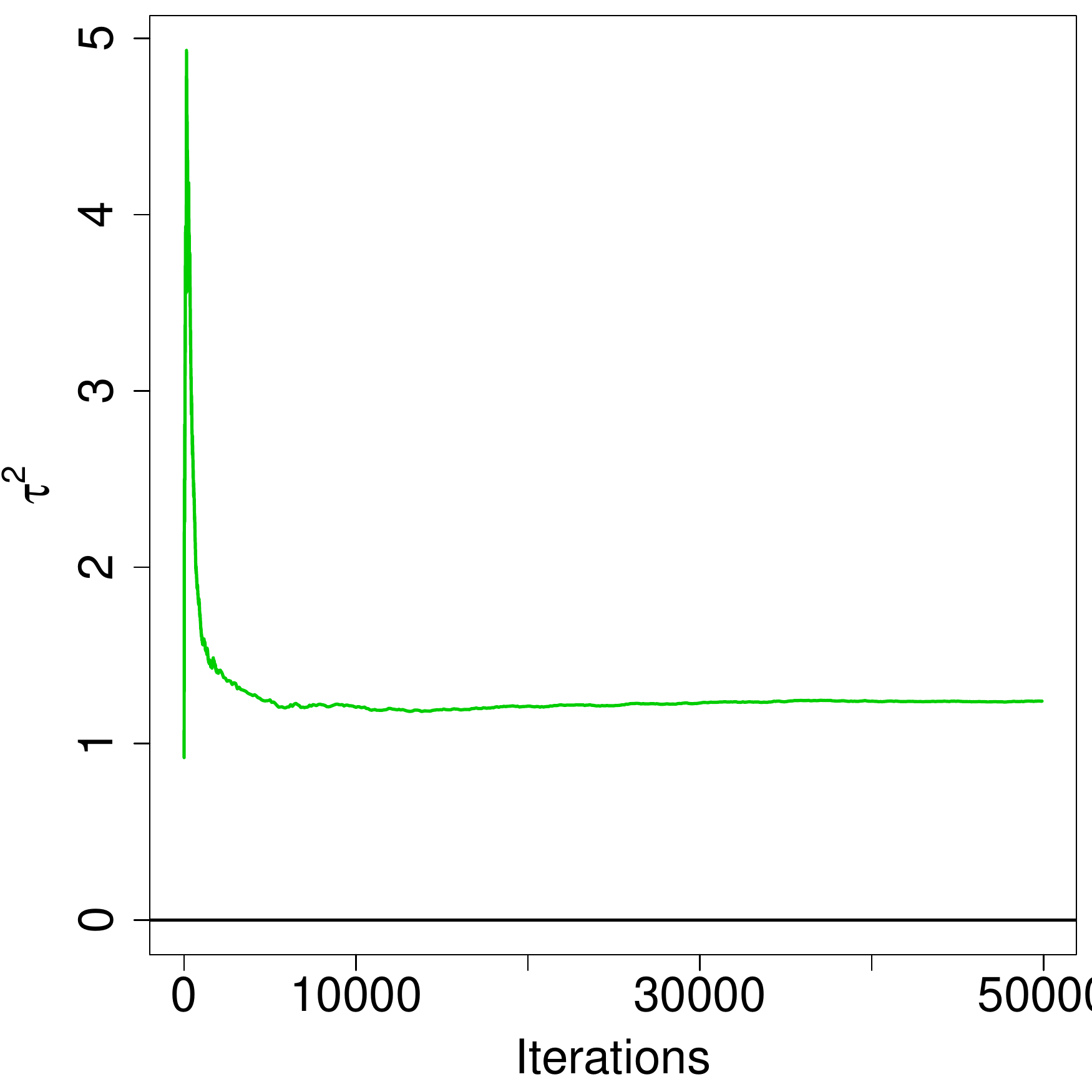}
\includegraphics[width=0.24\textwidth, page=2]{HT_n1500_prob095_Dof3_33333333333333_sigma1_diagnostics_new.pdf}
\includegraphics[width=0.24\textwidth, page=3]{HT_n1500_prob095_Dof3_33333333333333_sigma1_diagnostics_new.pdf}
\includegraphics[width=0.24\textwidth, page=4]{HT_n1500_prob095_Dof3_33333333333333_sigma1_diagnostics_new.pdf} \\
%
% INVERSE-GAMMA
%
\includegraphics[width=0.24\textwidth, page=1]{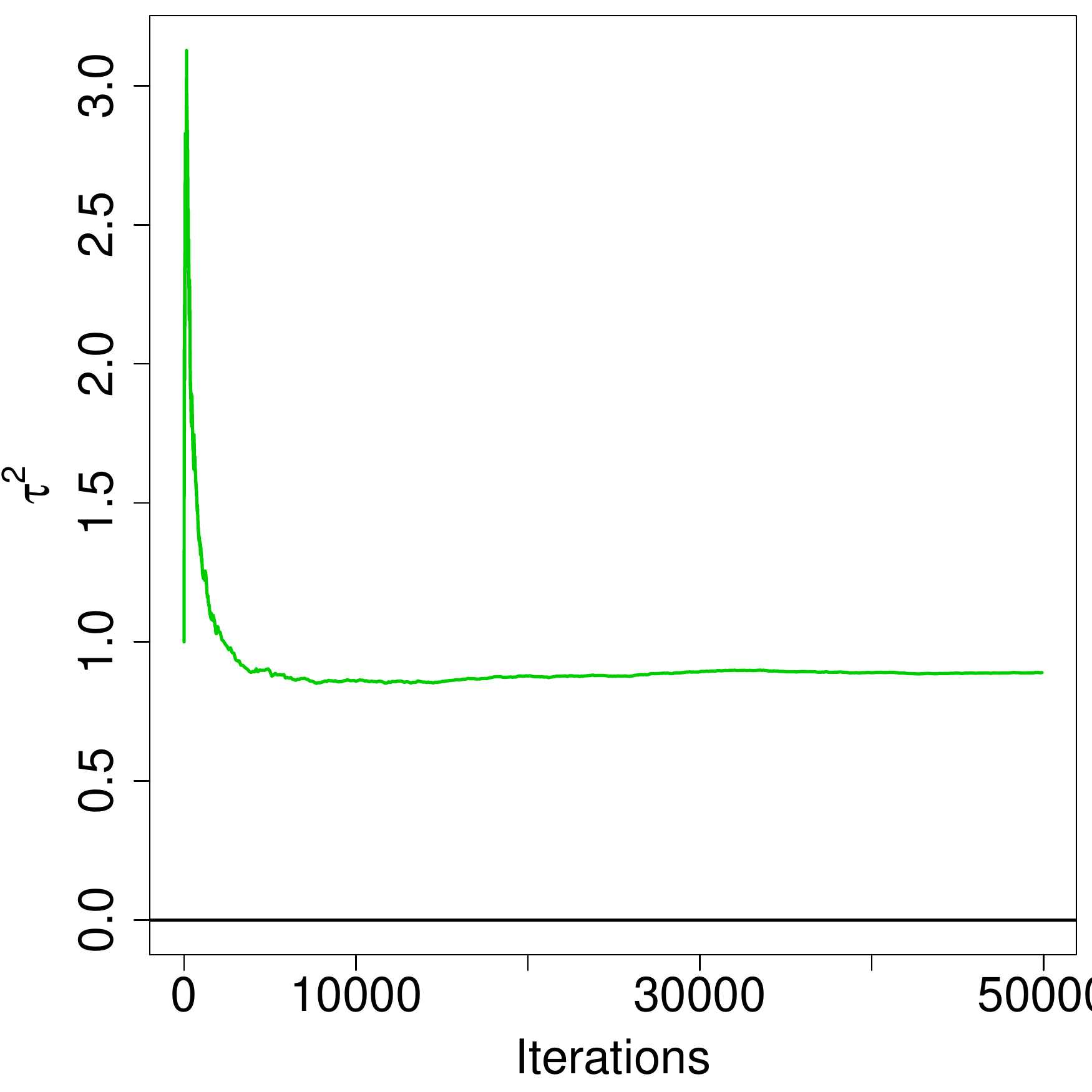}
\includegraphics[width=0.24\textwidth, page=2]{IG_n1500_prob095_scale10_shape5_diagnostics_new.pdf}
\includegraphics[width=0.24\textwidth, page=3]{IG_n1500_prob095_scale10_shape5_diagnostics_new.pdf}
\includegraphics[width=0.24\textwidth, page=4]{IG_n1500_prob095_scale10_shape5_diagnostics_new.pdf} \\
\caption{\small Univariate extreme quantile region results for $n=1,500$ Fr\'{e}chet (top row), Half-$t$ (middle) and inverse-Gamma (bottom) distributed data. Columns illustrate sampler scaling parameter $\tau^2$ (left) and overall acceptance probability against sampler iteration (centre-left) with target sampler acceptance rate of $\pi^*=0.234$ indicated by horizontal line. 
Centre-right column shows histogram and kernel density estimates of tail index $\gamma$  after removing $m=30,000$ iterations burn-in. Crosses are the lower and upper bounds of the estimated $95\%$ credibility interval.
Right column illustrates log-scale posterior densities of $p=1/750$ (light grey), $1/1500$ and $1/3000$ (dark grey) probability events.
Posterior mean and true tail index are indicated by dashed and solid lines respectively, observed data indicated by points on the $x$-axis.
}
\label{fig:univ_supp}
\end{figure}

We first replicate the simulation experiment in Section 4.1 of the main article, but this time censoring observations below the $95$-th empirical quantile. We simulate $n=1500$ observations 
from the distributions: $\mbox{Fr\'{e}chet}$ with location, scale and shape parameters equal to $\psi_0=3$, $\varsigma_0=1$ and $\xi_0=1/3$, respectively; 
$\mbox{Half-}t$ with scale and degrees of freedom equal to $\sigma_0=1$ and
$\nu_0=1/3$; $\mbox{Inv-Gamma}$ with shape parameters equal to $\eta_0=1/2$ and $\lambda_0=1$.
The results are illustrated in Figure~\ref{fig:univ_supp}, which is organised the same as Figure 1 in the main text.

Each row shows the results for each dataset: Fr\'{e}chet (top), Half-$t$ (middle) and inverse Gamma (bottom).
The left columns report the trace plots of the scaling parameter $\tau^2$ and the sampler average acceptance probability. The centre-right panels show histogram and kernel density estimates of the posterior distribution of the tail index $\gamma$, with dashed and solid vertical lines indicating the posterior mean and true value, respectively. Finally, the rightmost panels show the estimated posterior densities of quantiles corresponding to the exceedance probabilities $p=1/750$ (light grey), $1/1500$ and $1/3000$ (dark grey),  derived from (2.4$^{*}$). Vertical dashed and solid lines represent the posterior mean and true quantile values. 

The conclusions are similar to those obtained by censoring observations below the $90$-th empirical quantile; see the main text for details.
Once again, the true tail indices and quantiles always fall within the $95\%$ credible intervals obtained from the estimated posterior distributions. 
Using a higher threshold leads to the following differences with the results described in the main article: for the Fr\'{e}chet distribution (top row) the posterior means are all greater, and for the inverse-Gamma (bottom row) the posterior means are all greater. The posterior means are almost unchanged for the Half-$t$ distribution (middle row).
\\

\begin{figure}[t!]
\centering
%
% FRECHET
%
\includegraphics[width=0.25\textwidth, page=3]{FRE_n1500_prob09_loc30_scale10_shape3_3_diagnostics_new.pdf}
\includegraphics[width=0.25\textwidth, page=1]{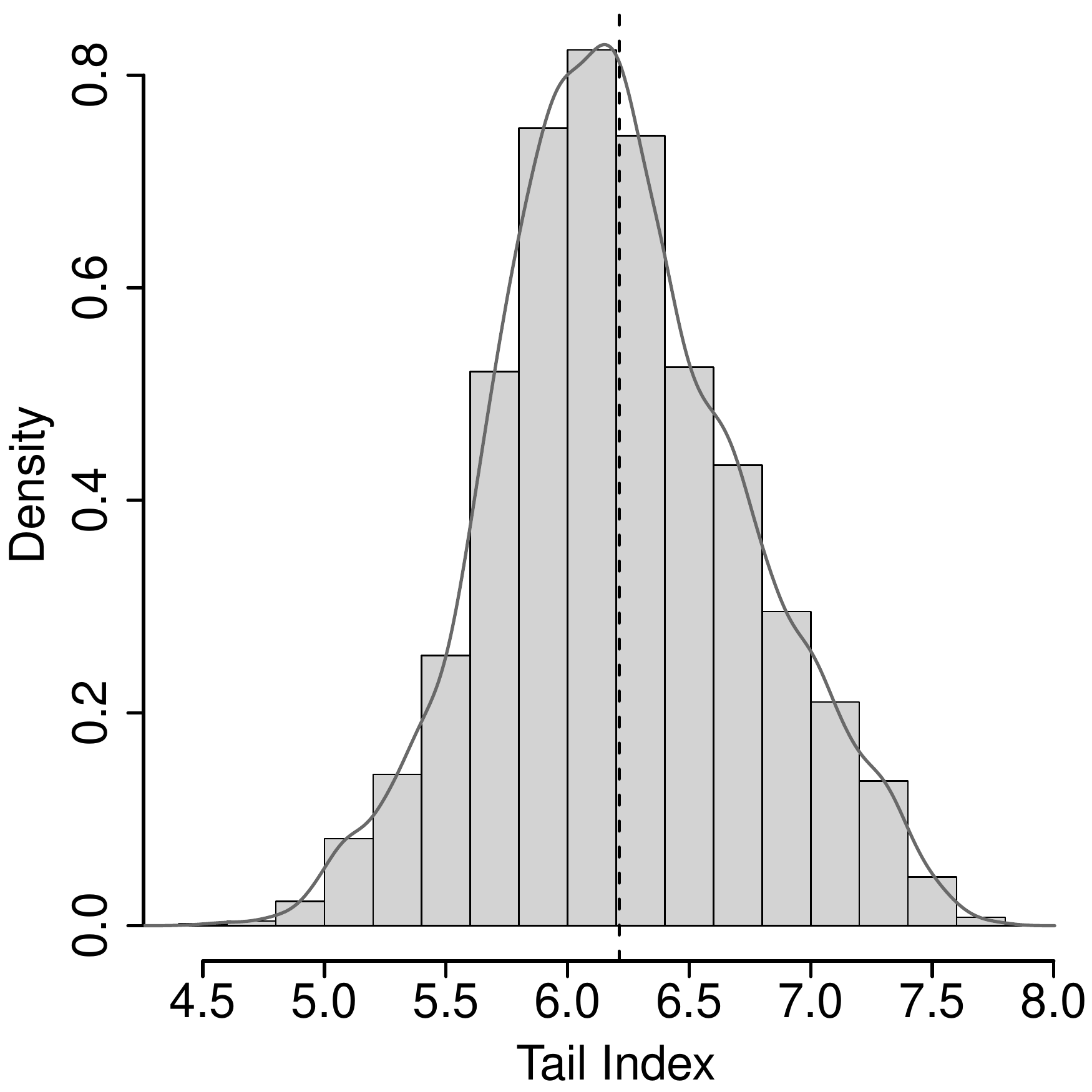}
\includegraphics[width=0.25\textwidth, page=1]{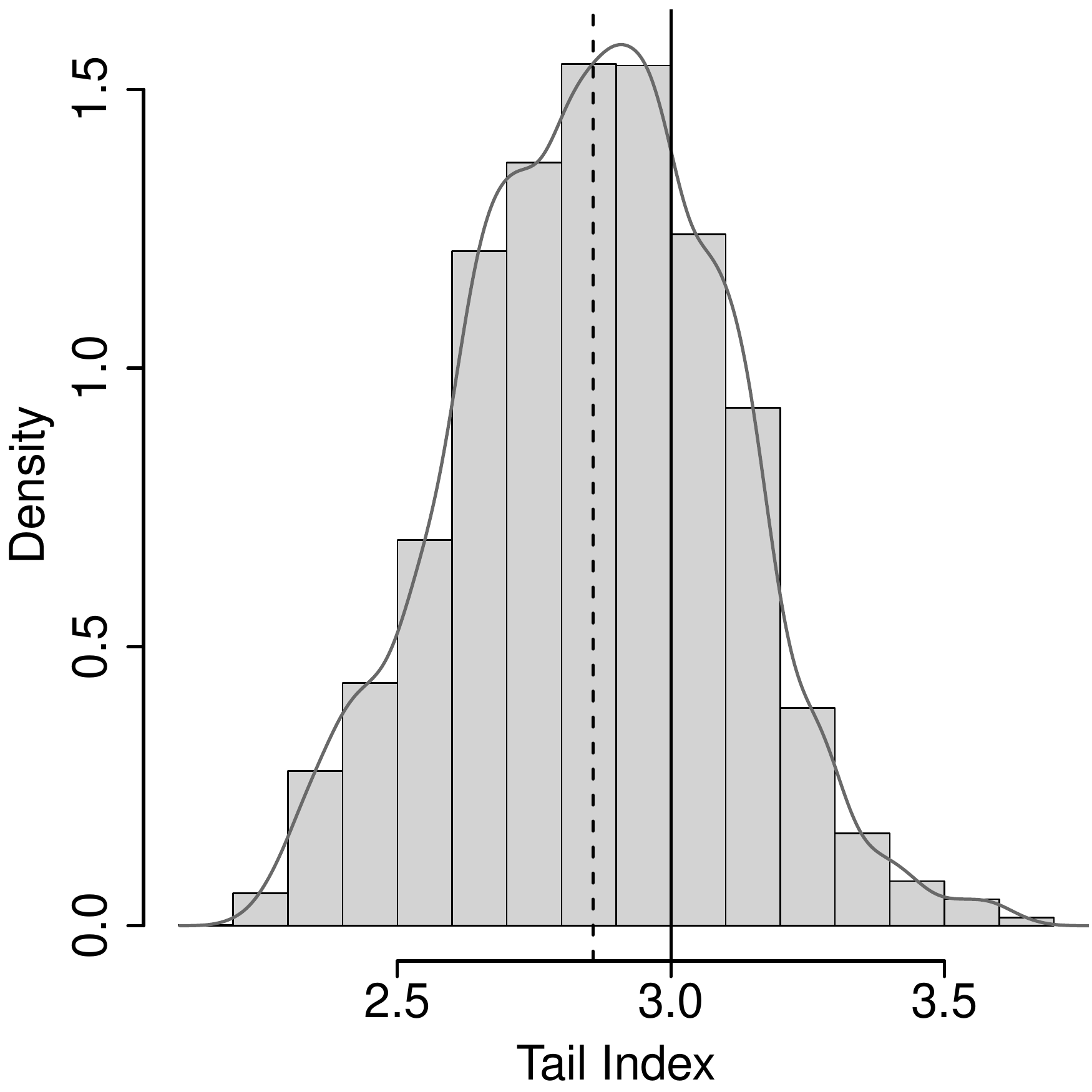} \\
\includegraphics[width=0.25\textwidth, page=4]{FRE_n1500_prob09_loc30_scale10_shape3_3_diagnostics_new.pdf}
\includegraphics[width=0.25\textwidth, page=2]{FRE_n1500_prob09_loc30_scale10_shape3_3_realgrey_scale2.pdf}
\includegraphics[width=0.25\textwidth, page=2]{FRE_n1500_prob09_loc30_scale10_shape3_3_realgrey_scale3.pdf}
\caption{\small Univariate extreme quantile region results for $n=1,500$ Fr\'{e}chet distributed data. Estimated posterior distributions obtained through the prior distributions A)-C) (from left to the right). 
Top row shows histogram and kernel density estimates of tail index $\gamma$  after removing $m=30,000$ iterations burn-in.
Bottom row illustrates log-scale estimated posterior densities of $p=1/750$ (light grey), $1/1500$ and $1/3000$ (dark grey) probability events.
Posterior mean and true tail index are indicated by dashed and solid lines respectively, observed data indicated by points on the $x$-axis.
}
\label{fig:sensitivity_frechet}
\end{figure}

\begin{figure}[t!]
\centering
%
% HALF-T
%
\includegraphics[width=0.25\textwidth, page=3]{HT_n1500_prob09_Dof3_33333333333333_sigma1_diagnostics_new.pdf}
\includegraphics[width=0.25\textwidth, page=1]{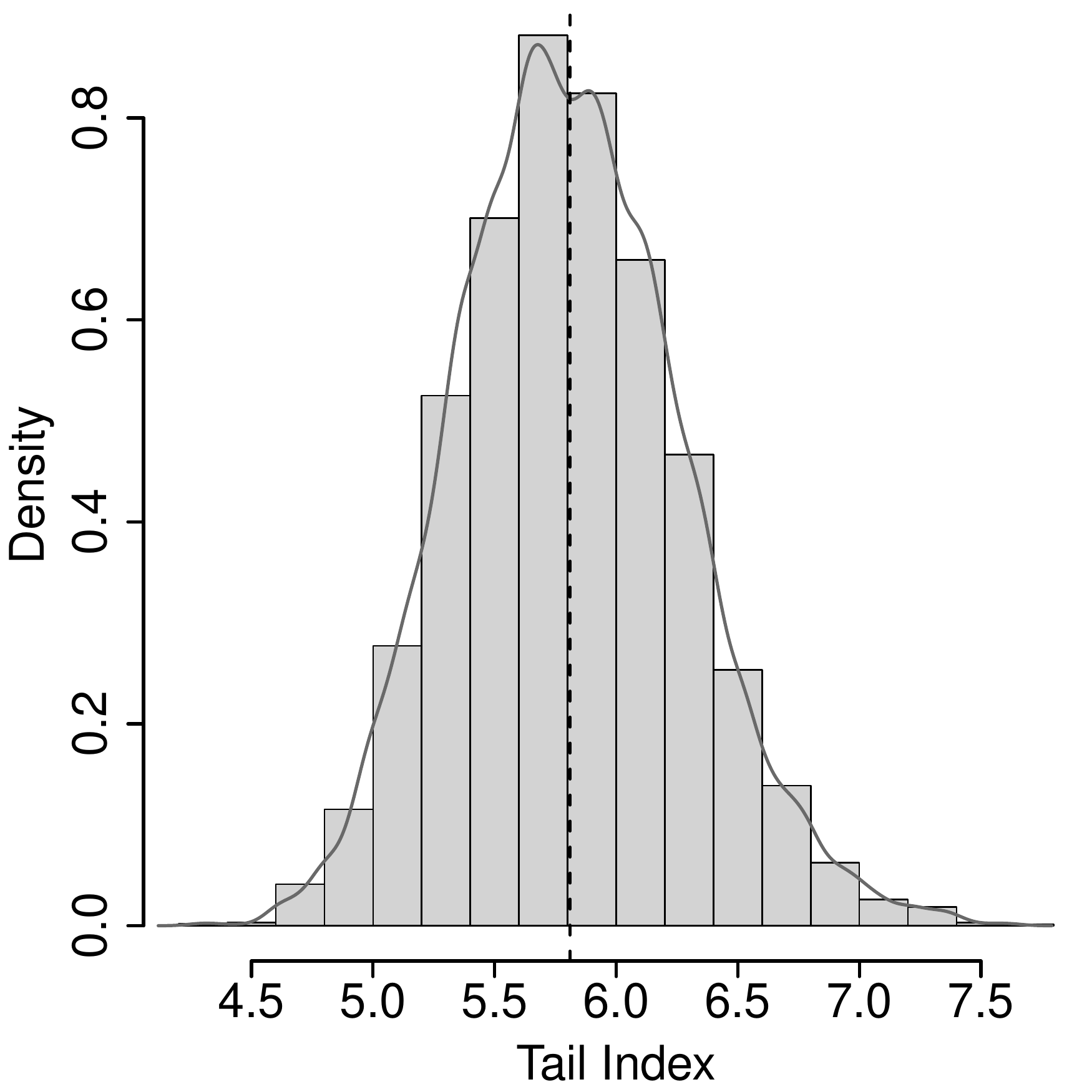}
\includegraphics[width=0.25\textwidth, page=1]{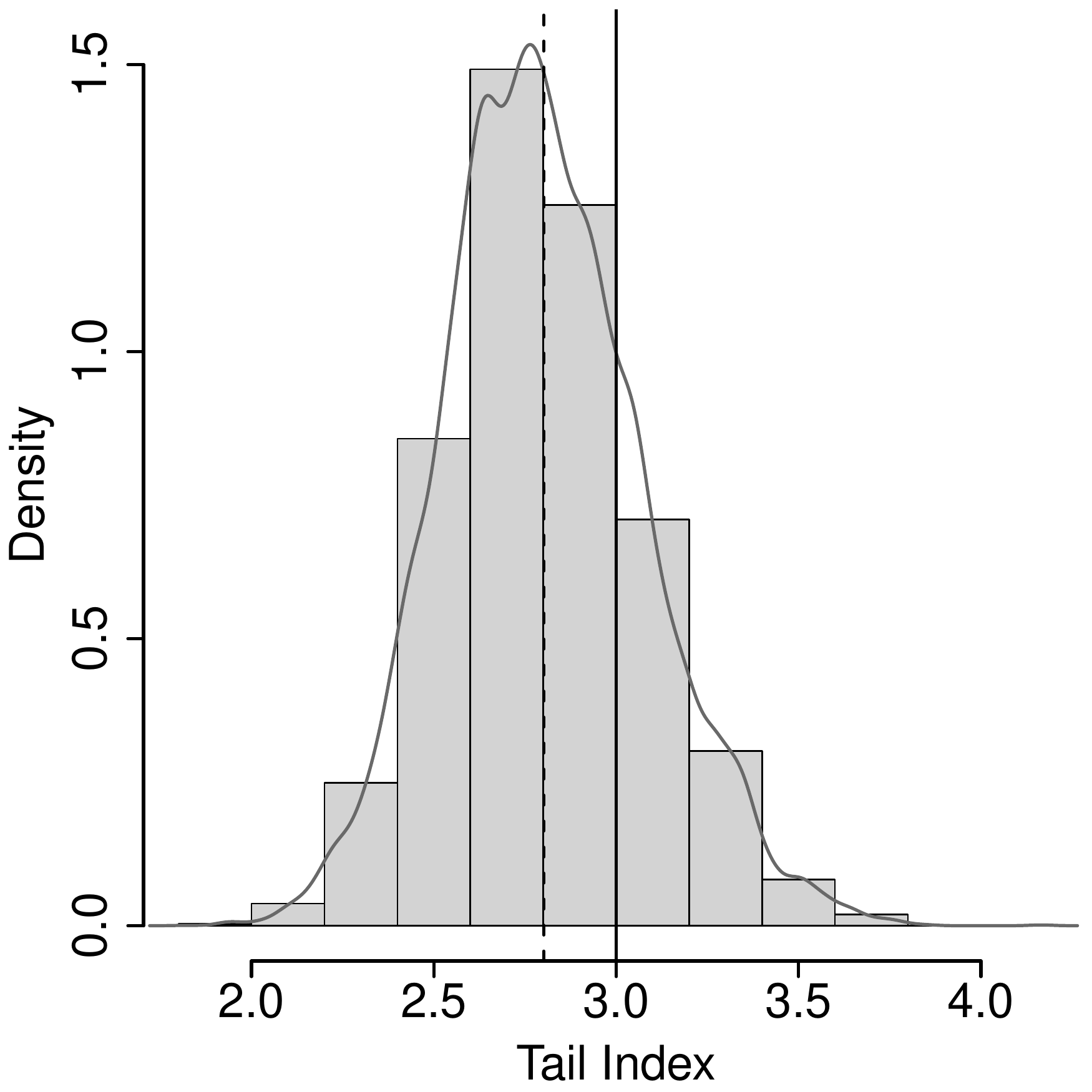} \\
\includegraphics[width=0.25\textwidth, page=4]{HT_n1500_prob09_Dof3_33333333333333_sigma1_diagnostics_new.pdf}
\includegraphics[width=0.25\textwidth, page=2]{HT_n1500_prob09_Dof3_33333333333333_sigma1_realgrey_scale2.pdf}
\includegraphics[width=0.25\textwidth, page=2]{HT_n1500_prob09_Dof3_33333333333333_sigma1_realgrey_scale3.pdf}
\caption{\small Univariate extreme quantile region results for $n=1,500$ Half-$t$ distributed data. This figure is organised as Figure \ref{fig:sensitivity_frechet}, see the latter for details.
}
\label{fig:sensitivity_half_t}
\end{figure}

\begin{figure}[h!]
\centering
%
% INVERSE-GAMMA
%
\includegraphics[width=0.25\textwidth, page=3]{IG_n1500_prob09_scale10_shape5_diagnostics_new.pdf}
\includegraphics[width=0.25\textwidth, page=1]{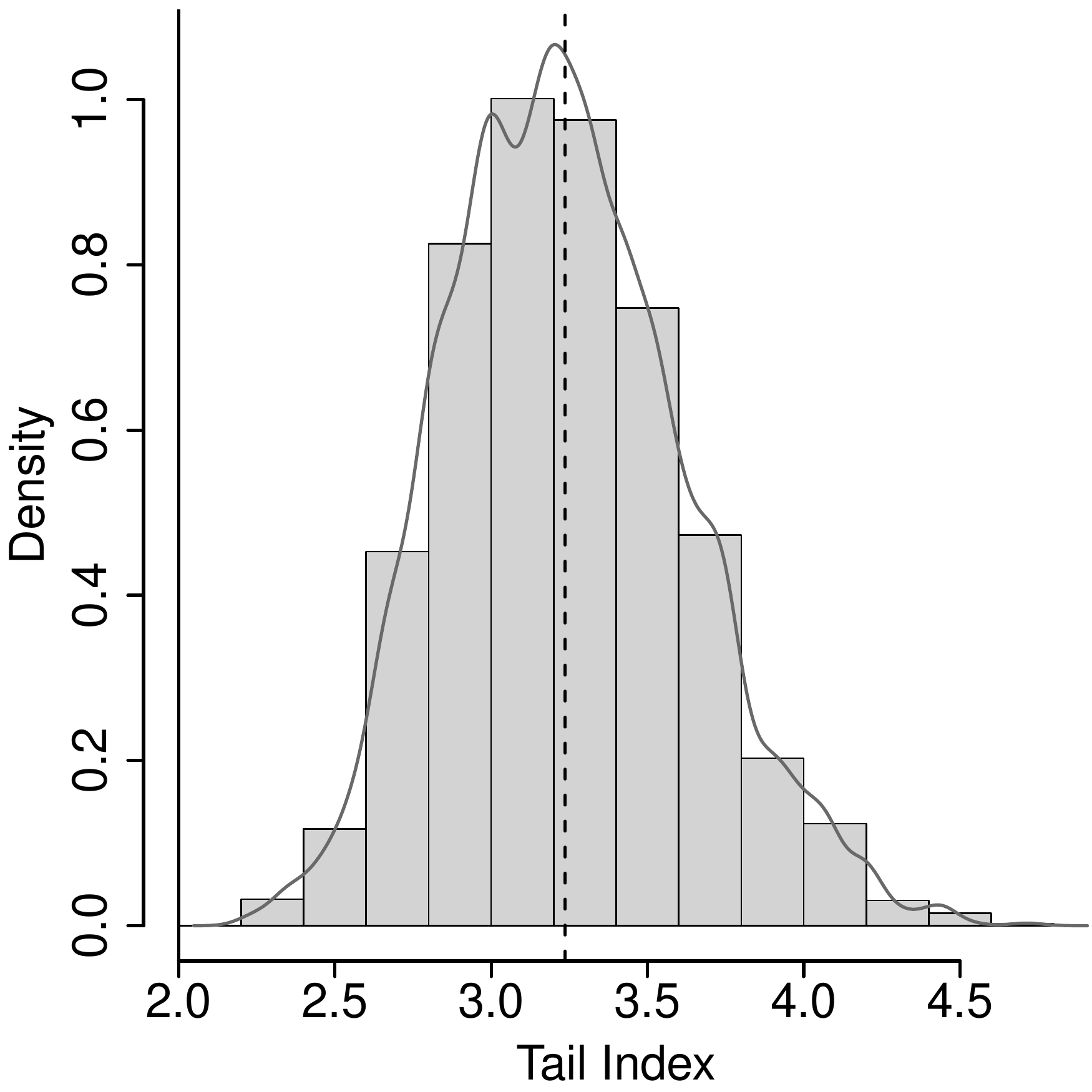}
\includegraphics[width=0.25\textwidth, page=1]{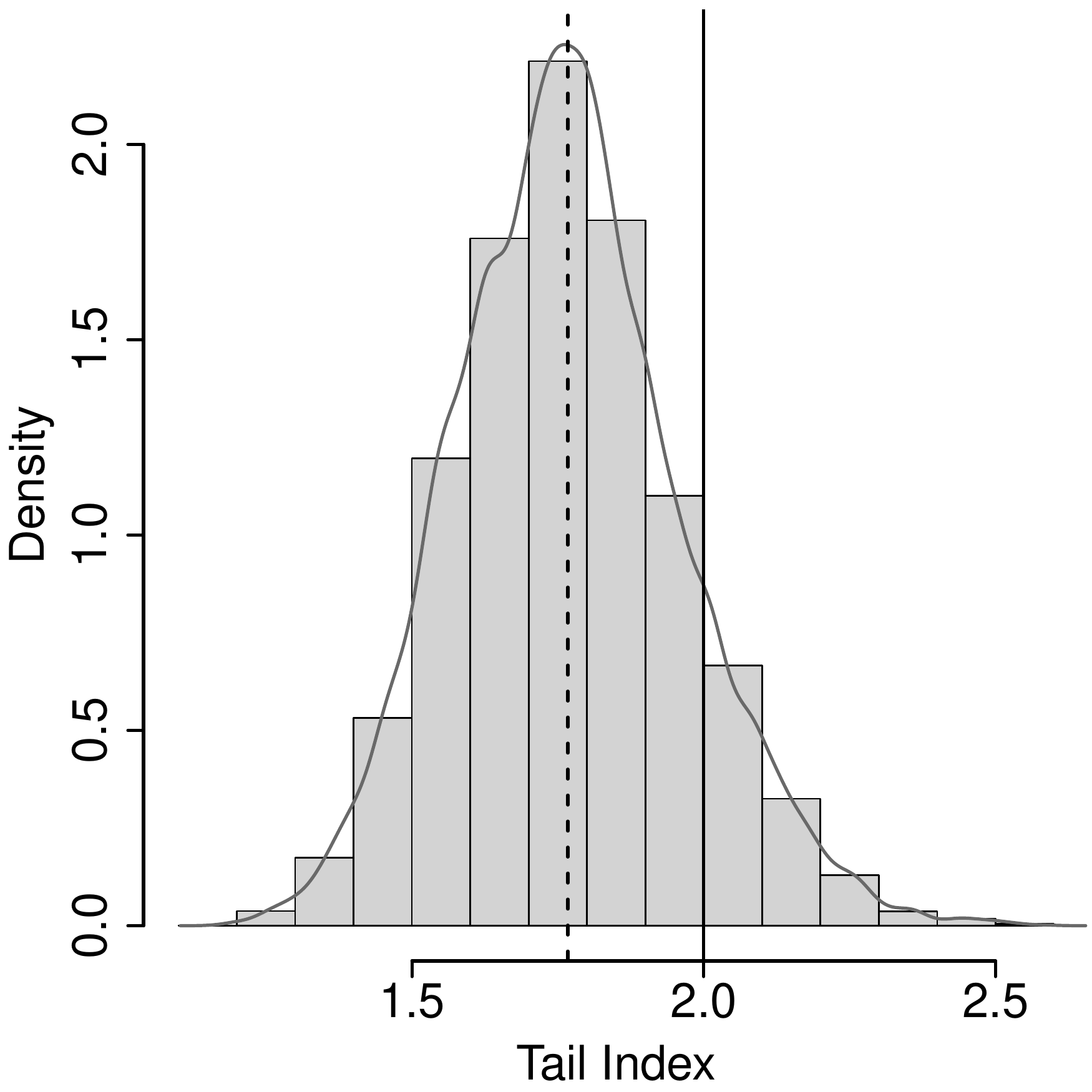} \\
\includegraphics[width=0.25\textwidth, page=4]{IG_n1500_prob09_scale10_shape5_diagnostics_new.pdf}
\includegraphics[width=0.25\textwidth, page=2]{IG_n1500_prob09_scale10_shape5_realgrey_scale2.pdf}
\includegraphics[width=0.25\textwidth, page=2]{IG_n1500_prob09_scale10_shape5_realgrey_scale3.pdf}
\caption{\small Univariate extreme quantile region results for $n=1,500$ inverse-Gamma distributed data. This figure is organised as Figure \ref{fig:sensitivity_frechet}, see the latter for details.
}
\label{fig:sensitivity_inv_gamma}
\end{figure}

We additionally perform a sensitivity analysis to examine the sensitivity of the posterior  to the prior distribution. 
For each univariate distribution considered in the main article (i.e.~$\mbox{Fr\'{e}chet}$, $\mbox{Half-}t$ and $\mbox{Inv-Gamma}$) we simulate $n=1500$ observations and then censor these below the $90$-th empirical quantile.
\noindent We consider the following three prior distributions:
\begin{itemize}
\item[A)] $\Pi(\mu,\sigma,\gamma):=\Pi(\mu)\cdot\Pi(\log(\sigma))\cdot\Pi(\gamma)$, where the prior distributions for $\mu$, $\log(\sigma)$ and $\gamma$ are uniform on the real line;
\item[B)] $\Pi(\mu,\sigma,\gamma): = \mathcal{N}\cdot(\mu;0,2^2)\cdot \mathrm{log}\mathcal{N}(\sigma|0, (1/3)^2)\cdot\mathcal{N}(\gamma;0,(3/2)^2)$, where the $\mathcal{N}(x;a,b^2)$ and $\mathrm{log}\mathcal{N}(x;a,b^2)$ denote normal and lognormal distribution with mean variance parameters $a$ and $b^2$, respectively;
\item[C)] $\Pi(\mu,\sigma,\gamma) := \mathcal{N}(\mu;0,5^2)\cdot \mathrm{log}\mathcal{N}(\sigma;0, 5^2)\cdot \mathcal{N}(\gamma;0,6^2)$.
\end{itemize}
Prior A) is the improper prior used in the main text, prior B) is strongly informative, and prior C) is proper but weakly informative.
Each MCMC sampler is then run for $M=50,000$ iterations for each simulated dataset.
The results are shown in Figures~\ref{fig:sensitivity_frechet}--\ref{fig:sensitivity_inv_gamma},
with columns (left-to-right) displaying the results under prior distributions A)--C).

The strongly informative prior distribution B) leads to estimated posterior distributions that are heavily influenced by the prior. As a result, the point and interval estimates of the tail index and quantiles are similarly driven by these beliefs, and so these do not correspond to the true values. Of course, if the practitioner truly believes this prior specification, then the resulting predictive inference is strictly correct in the usual Bayesian sense.

Under priors A) and C) the strength of prior information is much weaker, and as a result the posterior and resulting inferences are driven more by the data. The resulting point and interval estimates naturally accord more strongly with the true values in this case.

% =====================================================
% =====================================================
\section{Complement to Section~4.2}
% =====================================================
% =====================================================

We present two extra simulation experiments, involving estimating bivariate extreme quantile regions from the so-called Asymmetric and Clover densities.
We simulate $n=1500$ observations from each of the following bivariate densities:
 \begin{itemize}
\item {\bf Asymmetric distribution on $\R_+^2$:} The probability density of the asymmetric distribution is
$$
f(\bx) = \frac{c}{x_1^3 + x_2^4 +1}, \quad \bx \in \R_+^2,
$$
with $c \approx 0.58$. This density is heavy-tailed but its upper tail is less heavy than that of the bivariate Cauchy. Its tail indices  are $\gamma_1 = 4/5$ and $\gamma_2 = 3/5$, and the angular density is given by
$$
h(w) = \frac{6c}{25} \frac{c_1 c_2}{\left( (c_1 w^{\gamma_1})^3 + (c_2 (1-w)^{\gamma_2})^4 \right) w^{1-\gamma_1} (1-w)^{1-\gamma_2}},
$$
where $c_1 \approx 0.589$ and $c_2 \approx 0.593$. The angular basic density is
$$
q_*(w) = \left( \frac{c c_1 c_2}{(c_1 w^{\gamma_1})^3 + (c_2 (1-w)^{\gamma_2})^4} \right)^{-5/12},
$$
and the associated basic set  is
$$
\cS = \left\{ \bx\in\R^2_+: r > \left( \frac{c c_1 c_2}{(c_1 w^{\gamma_1})^3 + (c_2 (1-w)^{\gamma_2})^4} \right)^{5/12},\, w \in [0, 1] \right\}.
$$
\item {\bf Clover distribution on $\R_+^2$:} The probability density of the Clover distribution
is 
$$
f(\bx) = \frac{64}{25 \pi} \frac{\left( x_1^2 + \left( (x_2+1)^{4/5} -1 \right)^2 \right)^2 - 3x^2 \left( (x_2+1)^{4/5} -1 \right)^2}{(x_2+1)^{1/5} \left( 1 + x_1^2 + \left( (x_2+1)^{4/5} -1 \right)^2 \right)^{3/2} \left( x_1^2 + \left( (x_2+1)^{4/5} -1 \right)^2 \right)^2}.
$$
This density is heavy-tailed and its joint upper tail is heavier than that of the the bivariate Cauchy. Its tail indices are $\gamma_1 = 1$ and $\gamma_2 = 5/4$, and the angular density is
$$
h(w) = \frac{4c \left( w^4 - w^2(1-w)^2 + (1-w)^4 \right)}{\left(w^2 + (1-w)^2 \right)^{7/2}},
$$
where $c \approx 0.208$. The angular basic density is
$$
q_*(w) = \left( \frac{32c}{5} \frac{w^4 - w^2(1-w)^2 + (1-w)^4}{\left(w^2 + (1-w)^2 \right)^{7/2} \left( 1-w \right)^{1/4}} \right)^{-4/13},
$$
and the associated basic set is
$$
\cS = \left\{\bx\in\R^2_+: r >  \left( \frac{32c}{5} \frac{w^4 - w^2(1-w)^2 + (1-w)^4}{\left(w^2 + (1-w)^2 \right)^{7/2} \left( 1-w \right)^{1/4}} \right)^{4/13}, 
w \in [0, 1] \right\}.
$$
\end{itemize}
\begin{figure}[t]
\includegraphics[width=\textwidth]{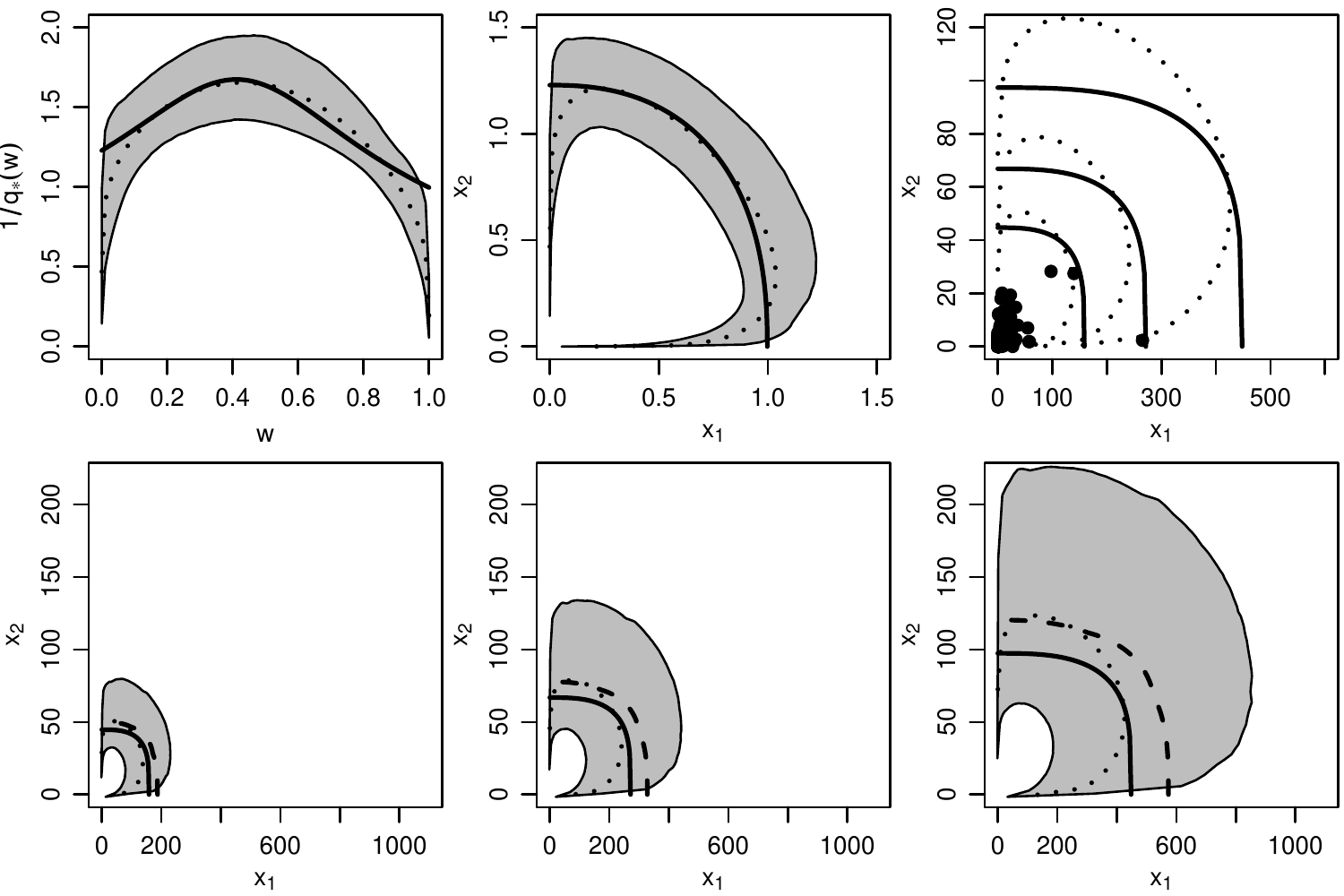}
\caption{\small Extreme quantile analysis for the asymmetric distribution. True (solid lines), posterior mean estimate (dotted) and $90\%$ credible regions (grey) for the inverse of the angular basic density (top left), the basic set $\cS$ (top middle), and the quantiles with probability $p=1/750, 1/1500$ and $1/3000$ (bottom left to bottom right). Dashed lines in bottom panels shows the EdHK point estimator. Top right panel illustrates the simulated dataset (points) and true and posterior mean estimated quantiles. }
\label{fig:Asymmetric}
\end{figure}
\begin{figure}[t]
\includegraphics[width=\textwidth]{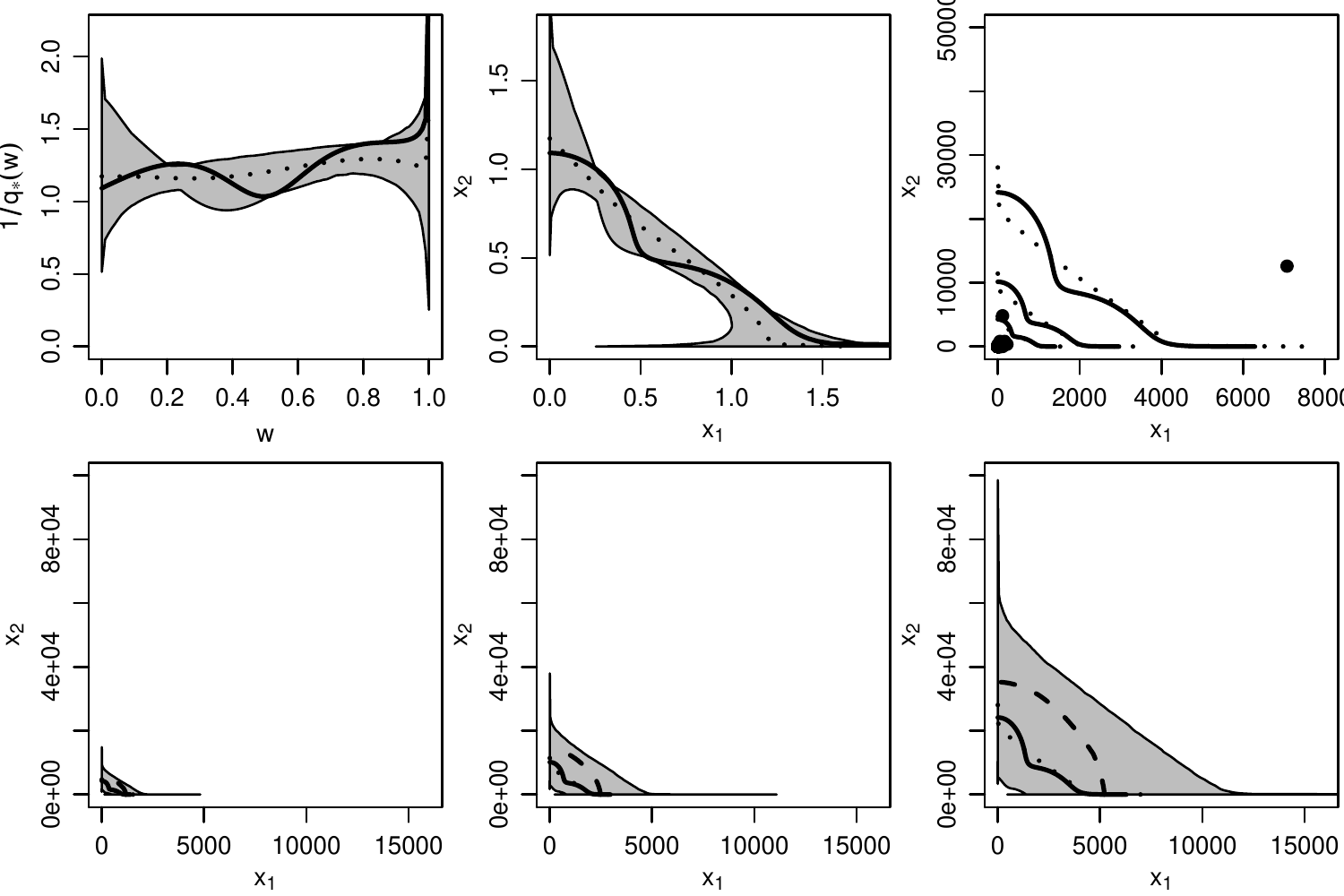}
\caption{\small 
Extreme quantile analysis for the clover distribution. True (solid lines), posterior mean estimate (dotted) and $90\%$ credible regions (grey) for the inverse of the angular basic density (top left), the basic set $\cS$ (top middle), and the quantiles with probability $p=1/750, 1/1500$ and $1/3000$ (bottom left to bottom right). Dashed lines in bottom panels shows the EdHK point estimator. Top right panel illustrates the simulated dataset (points) and true and posterior mean estimated quantiles.}
\label{fig:Clover}
\end{figure}

For each margin we censor all observations that fall below the corresponding $90$-th marginal empirical quantile. We specify the prior distribution $\Pi(\mu_i,\sigma_i,\gamma_i)\propto1/\sigma$ for each margin $i=1,2$ and a $\Pi(\kappa)=\mathrm{NegBin}(m_{NB}=3.2, \sigma_{NB}=4.48)$ prior distribution for the polynomial degree. We allow for the possibility of non-zero point masses at the endpoints of the simplex by specifying the uniform prior distributions $\Pi(p_0) = \Pi(p_1) = \textrm{Unif}(0, 0.1)$.
We run Algorithm 1 in the main text for $M=50,000$ iterations and determine the burn-in period $m$ by visual inspection of trace plots of the marginal scaling parameters $\tau_1, \tau_2$ and the overall acceptance rates of the marginal ($\pi_i^*=0.234, i=1,2$) and dependence proposals. Then, we estimate the bivariate quantile regions corresponding to the exceedance probabilities $p=1/750$, $1/1500$ and $1/3000$. 

The complete description on how to derive the estimated posterior densities for the angular basic density $q_*(w)$, the basic set $\cS$ and the quantile $\widetilde{Q}_n$ and their respective posterior means and credibility bands is given in Section~4.2 of the main text.

The results for the asymmetric distribution (Figure~\ref{fig:Asymmetric}) indicate a generally well captured dependence structure (top-left panel), albeit with an estimated dip down towards $0$ at the boundaries of the simplex. This corresponds to a basic set where values of one component that are close to $0$ also tend to drag the other component towards $0$. The top right panel illustrates that the quantile regions follow the behaviour of the data quite well, particularly the largest observations.
The mean quantile estimates (bottom panels) appear to closely match the true levels, while the EdHK estimates do not capture either the shape or the magnitude of the region well, and do not convincingly lie within the 90\% credible quantile regions. 

The structure of the clover distribution (Figure~\ref{fig:Clover}) is more challenging to model. Our procedure, despite having some difficulty capturing the detailed fluctuations for the given observed dataset size ($n=1500$), appears to estimate the general trend of the dependence behaviour fairly well. 
Indeed, the inverse of the true angular basic density $q_*^{-1}(w)$ and the true basic set $\cS$ are almost fully contained within the relatively narrow 90\% credible regions. 
The EdHK estimator performs reasonably poorly, giving higher quantile estimates than our mean quantile estimate, with both overestimating the true quantile level, although the overestimation is only slight for our mean estimator. This can be explained by the presence of an observation above the $(2999/3000)$ quantile and two observations above the $(1499/1500)$ quantile.

\end{document}